\documentclass[twocolumn,showpacs,nofootinbib,aps,superscriptaddress,eqsecnum,prd,notitlepage,showkeys,10pt]{revtex4-1}
\usepackage{amssymb}
\usepackage{amsmath}
\usepackage{physics}
\usepackage{graphicx}
\usepackage{dcolumn}
\usepackage{hyperref}
\usepackage{xcolor}
\usepackage[normalem]{ulem}
\usepackage{siunitx}

\DeclareMathAlphabet\mathbfcal{OMS}{cmsy}{b}{n}

\definecolor{darkorange}{rgb}{1.0, 0.55, 0.0}
\definecolor{darkgreen}{rgb}{0.0, 0.5, 0.0}

\begin{document}

\title{Fast Pulses for High-Fidelity Circularization of Interacting Rydberg atoms}

\author{Matthias Hüls}
\affiliation{Forschungszentrum Jülich GmbH, Peter Grünberg Institute, Quantum Control (PGI-8), 52425 Jülich, Germany}
\affiliation{Institute for Theoretical Physics, University of Cologne, 50937 Köln, Germany}

\author{Aurore A. Young}
\affiliation{Laboratoire Kastler Brossel, Collège de France, CNRS, ENS-Université PSL, Sorbonne Université, 11, place Marcelin Berthelot, 75005 Paris, France}

\author{Clément Sayrin}
\affiliation{Laboratoire Kastler Brossel, Collège de France, CNRS, ENS-Université PSL, Sorbonne Université, 11, place Marcelin Berthelot, 75005 Paris, France}
\affiliation{Institut Universitaire de France, 1 rue Descartes, 75231 Paris Cedex 05, France}

\author{Michel Brune}
\affiliation{Laboratoire Kastler Brossel, Collège de France, CNRS, ENS-Université PSL, Sorbonne Université, 11, place Marcelin Berthelot, 75005 Paris, France}

\author{Jean-Michel Raimond}
\affiliation{Laboratoire Kastler Brossel, Collège de France, CNRS, ENS-Université PSL, Sorbonne Université, 11, place Marcelin Berthelot, 75005 Paris, France}

\author{Tommaso Calarco}
\affiliation{Forschungszentrum Jülich GmbH, Peter Grünberg Institute, Quantum Control (PGI-8), 52425 Jülich, Germany}
\affiliation{Institute for Theoretical Physics, University of Cologne, 50937 Köln, Germany}
\affiliation{Dipartimento di Fisica e Astronomia, Università di Bologna, 40127 Bologna, Italy}

\author{Felix Motzoi}
\affiliation{Forschungszentrum Jülich GmbH, Peter Grünberg Institute, Quantum Control (PGI-8), 52425 Jülich, Germany}
\affiliation{Institute for Theoretical Physics, University of Cologne, 50937 Köln, Germany}

\author{Robert Zeier}
\affiliation{Forschungszentrum Jülich GmbH, Peter Grünberg Institute, Quantum Control (PGI-8), 52425 Jülich, Germany}

\author{Eloisa Cuestas}
\affiliation{Forschungszentrum Jülich GmbH, Peter Grünberg Institute, Quantum Control (PGI-8), 52425 Jülich, Germany}
\affiliation{OIST Graduate University, Onna, Okinawa, Japan}

\begin{abstract}
Circular states in Rydberg atoms offer a promising platform for quantum computation, quantum simulation and quantum sensing. However, the final step of their preparation ---termed as \textit{circularization}, a process that involves the transfer of a large amount of angular momentum quanta to the valence electron by means of radio-frequency (RF) pulses--- remains as a major bottleneck for all technological applications based on interacting circular Rydberg atoms. Even though successfully implemented to circularize an atom cloud in the dilute regime, previous efforts to speed up the circularization process have focused on the single-atom case, thereby neglecting the interactions which constitute one of the main resources for quantum simulation and computation. In this theoretical work we show how interactions between two atoms disturb the efficiency of pulses designed for single atoms and identify shifts induced by the interactions on relevant transition energies as the dominant disturbance. We demonstrate that the initial efficiency of single-atom pulses can be restored by adapting them to these shifts. Our approach is based on a simple functional form depending only on two linear parameters, which we derive analytically. The adapted pulses prepare two $^{87}$Rb atoms after $\SI{65}{\nano\second}$ in a $n=52$ circular state with a fidelity of at least $\SI{95}{\percent}$ for interatomic distances down to $\SI{6.5}{\micro\meter}$ and for all angular configurations, while also complying experimental amplitude and frequency constraints. Finally, we show that when combining our adapted pulses with Krotov's pulse-shaping algorithm we obtain high-fidelity pulses for any pair arrangement with interatomic distances larger than $\SI{5.9}{\micro\meter}$. This work demonstrates that fast RF pulses can circularize interacting Rydberg atoms, paving the way toward their technological application.
\end{abstract}


\maketitle


\section{Introduction}
\label{sec_intro}

Circular Rydberg states are characterized by maximal angular momentum $l=n-1$ and $|m|=n-1$ within a state manifold of high principal quantum number. They exhibit several properties that make them promising for multiple quantum technologies. Their long lifetimes \cite{Hlzl2024, Wu2023}, together with strong longe-range dipole-dipole interactions \cite{Ravon2023} turn them into ideal candidates for quantum simulation and computation \cite{Nguyen2018,Cohen2021, Meinert2020}. In the spirit of quantum simulation, the dipole-dipole interactions between atoms in circular states of different $n$ prepared in a chain mimic an XXZ Hamiltonian of a spin-1/2 system with nearest-neighbour interactions \cite{Nguyen2018}, while arrays of interacting circular Rydberg atoms have been proposed as a quantum computing platform that could achieve two-qubit gate errors of the order of $10^{-5}$ \cite{Cohen2021}. Moreover, due to their strong interactions with electromagnetic fields, they are of practical interest for quantum sensing; Schrödinger-cat states based on circular states have enabled highly sensitive measurements of electric and magnetic fields \cite{Facon2016, Dietsche2019}. 

The final step of the experimental preparation of circular Rydberg atoms (or \textit{circularization}) involves the transfer of a large amount of angular momentum quanta to the single valence electron and appears as the major bottleneck for all of their technological applications. In particular, it is highly desirable to minimize the time devoted for this preparation such that the available time for the realization of quantum gates, running the quantum simulator, or performing electrometry or magnetometry experiments can be maximized. In current experimental implementations, the transfer of population from an initial low-$m$ Rydberg state towards the circular state is usually adiabatically driven by coupling the atom to a $\sigma_+$ polarized radio-frequency (RF) field and slowly ramping internal state transitions through resonance using an electric field \cite{Hlzl2024, Ravon2023, Nussenzveig1993}. During this process the atom absorbs the necessary angular momentum by means of the RF photons and reach the maximal angular momentum state. This method, called \textit{adiabatic passage}, circularizes non-interacting atoms with an estimated fidelity of $\SI{99.5}{\percent}$ \cite{Larrouy2020}, but because of the adiabaticity condition it comes at the cost of requiring a total time of roughly $\SI{4}{\micro\second}$ for the $n=52$ states of $^{87}$Rb. In atomic arrays, interatomic interactions reduce the efficiency of this method. Recent experiments have therefore relied on either arranging the atoms at large interatomic distances or at a specific \textit{magic angle} relative to the external fields defining the quantization axis, where the interactions are sufficiently weak to be neglected \cite{Ravon2023, Mhaignerie2025}. Even though beneficial for the maximization of the final circular state probability, this intentional suppression of interactions must be overcome for quantum simulation and computation applications where the interactions between Rydberg atoms play a crucial role \cite{Browaeys2020, Shi_2022, saffman_2010}. Importantly, the circularization strategy based on the magic angle is fundamentally limited to one-dimensional chains and cannot be used to create two-dimensional arrays of circular atoms or more complex geometries.

In this work, we address the need of developing RF pulses for a fast and efficient circularization of Rydberg atoms in the presence of interactions. We present a general strategy for the accurate and simultaneous preparation of two interacting atoms in the same circular Rydberg state without relying on adiabaticity or particular atom arrangements. Our optimization scheme hinges on an analytical adaptation of the pulses developed for the circularization of a single-atom (see Refs. \cite{Patsch2018,Larrouy2020}) which leads to a circular pair state probability higher than $\SI{95}{\percent}$ for two $^{87}$Rb interacting atoms in the $n=52$ Rydberg manifold with interatomic distances down to $\SI{6.5}{\micro\meter}$ and for all angular configurations while at the same time complying with amplitude and bandwidth experimental constraints. Furthermore, by combining the adapted pulse with Krotov's pulse shaping algorithm \cite{Krotov1983, Sklarz2002, Reich2012}, we obtain high-fidelity pulses for any pair arrangement. The obtained pulses speed up the circularization in the presence of interactions by a factor of $60$ with respect to the adiabatic passage, therefore reducing the effect of relaxations induced by the interactions to a neglibile amount.

This paper is organized as follows. In Sec. \ref{sec_single_atom} we review the main concepts for the analysis of the circularization process by shaping pulses for a single rubidium atom with a static magnetic field as the only new ingredient with respect to the situation considered in Ref. \cite{Patsch2018}. In Sec. \ref{sec_atom_pair} we focus on an interacting pair of atoms and show that the pulse optimized for the circularization of a single atom fails in the presence of interactions. In Sec. \ref{sec_frequency_shifts} we build a physical picture based on conserved quantities that allows us to identify the main mechanism through which the interactions disturb the circularization. In Sec. \ref{sec_ad_freq} we present a pulse shaping strategy for accurately circularizing the interacting atoms. We provide a simple functional form depending only on two parameters to adapt the pulses optimized for a single atom to the interacting case. Furthermore, we give explicit analytical formulas obtained in the hydrogenic picture for the parameters designed to counteract the effects of interactions. In Sec. \ref{sec_optimization} we propose a multi-layer scheme to speed up the optimization of the circularization task. We summarize our findings and conclude in Sec.~\ref{sec_conclus}. Details of the calculations and simulations are given in Appendices \ref{Appendix: Details on simulation} to \ref{Appendix: state space reduction}.

\section{Circularization of a single Rydberg atom}
\label{sec_single_atom}

\begin{figure*}[t]
 \centering \includegraphics[width=\textwidth]{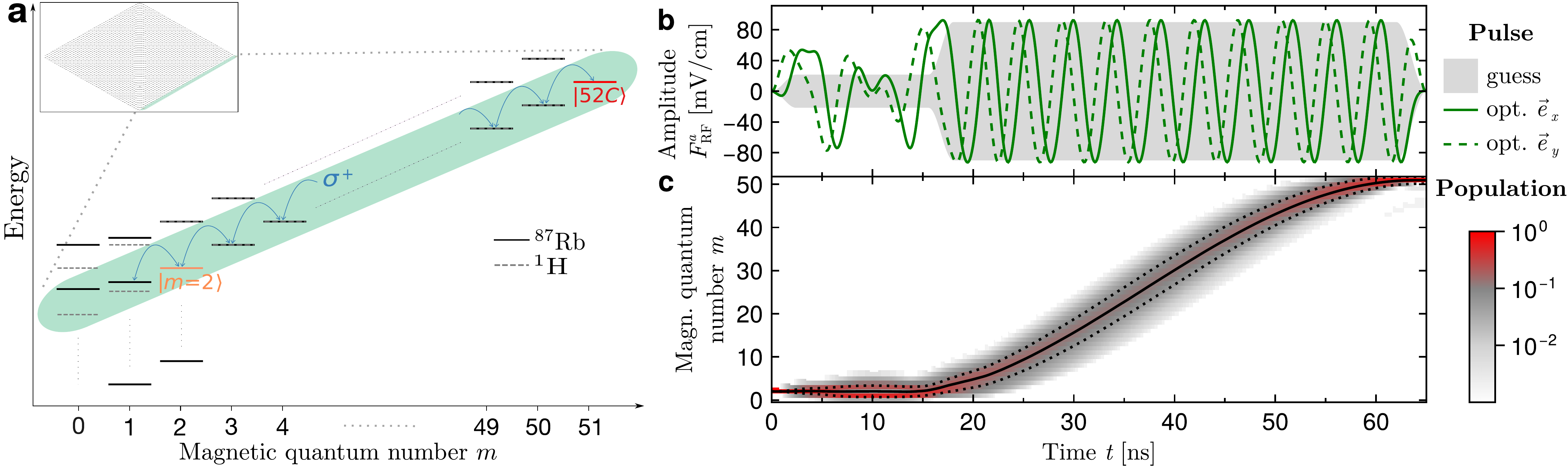}
   \caption{(a) Rydberg levels of the $n = 52$ manifold of $^1$H (dashed gray lines) and $^{87}$Rb (solid black lines) in the presence of static electric and magnetic fields of magnitude $\mathcal{E}= \SI{2.1}{\volt\per\centi\meter}$ and $\text{B} = 14 \text{ G}$. The eigenenergies ordered by their corresponding magnetic quantum number $m$ reveal the diamond-like pattern depicted in the inset. For hydrogen the subspace of the energetically lowest states with $m\geq 0$ forms the $\textit{lowest diagonal ladder}$ (shaded turquoise) that includes the circular state $\ket{52C}$ (red).
   The lowest diagonal ladder is harmonic up to the dominant first order Stark shift with transitions $m \to m+1$ (blue arrows) of $\SI{229.2}{\mega\hertz}$ usually driven by a $\sigma_+$ polarized RF field. The $m=0$ and $m=1$ levels of rubidium strongly affected by quantum defects lie outside of the hydrogenic-like manifold. Since the transition from $m=2$ to $m=3$ is nearly resonant with the hydrogenic ladder, we choose $\ket{m=2}$ (orange) as the initial state for rubidium. While the figure depicts two diagonal ladders, in our simulations we included all the states in the five lowest diagonal ladders. (b) $x$- and $y$-component of the RF pulse $\mathbf{F}_{\text{opt}}^{\text{a}}$ for the circularization of a rubidium atom in the $n=52$ Rydberg manifold (solid and dashed lines, respectively) obtained after a full optimization of a $\sigma_+$ polarized guess pulse (gray envelope) for a duration of $\SI{65}{\nano\second}$. (c) Evolution of the population (color scale) in the lowest diagonal ladder under the action of $\mathbf{F}_{\text{opt}}^{\text{a}}$. The expected value of the projection of the angular momentum $\bar{m}$ (solid black curve) increases monotonically until reaching the circular state, its uncertainty $\sigma_m$ (standard deviation, dotted lines) remains under 4 for all the time evolution. The optimized pulse transfers the initial state $\ket{m = 2}$ to the circular state $\ket{m = 51}$ with the targeted probability of $\SI{99}{\percent}$.}
 \label{fig_single_atom}
\end{figure*}

We begin by introducing the principles of a circularization with fast RF pulses for hydrogen and subsequently extend the discussion to rubidium, the atomic species used in the experiment \cite{Larrouy2020, Patsch2018, Signoles2017}. We focus on the preparation of the circular state of the $n=52$ state manifold, from now on denoted as $\ket{52\,\text{C}}$. To access all the states in a fixed-$n$ state manifold, degeneracies are usually lifted with a static electric field $\mathbfcal{E}$ that defines the $z$-direction. We set the latter to a magnitude of $\mathcal{E} = \SI{2.1}{\volt\per\centi\meter}$. Inside a fixed-$n$ manifold, the eigenstates of the free Hamiltonian $\hat{h}_0 $ and electric field Hamiltonian $\hat{h}_{\mathcal{E}}$ can be approximated by what is known as the \textit{parabolic basis} $\{\ket{n, k, m}\}$, where $n$ and $m$ are the usual principal and magnetic quantum numbers while $k$ is the \textit{eccentricity quantum number} that is proportional to the projection of the dipole moment $\hat{\mathbf{d}}$ on the $z$-axis \cite{Jannussis1979, FernndezMenchero2013}. When ordering the states $\ket{n, k, m}$ by eigenenergy and $m$, a diamond-like shape is revealed: the \textit{Stark diamond} having the circular state $\ket{52\,\text{C}} = \ket{n = 52, k = 0, m = 51}$ on its outer right tip, see Fig.~\ref{fig_single_atom}(a). The circular state $\ket{52\,\text{C}}$ is also part of the \textit{lowest diagonal ladder} indicated by the turquoise shade in Fig.~\ref{fig_single_atom}(a), which is the subspace $\{ \ket{m} = \ket{n, k = -(n-1-m), m} \}$ of the energetically lowest states with $0 \leq m \leq n-1$, forming a ladder that is harmonic up to linear Stark shifts. For our settings, the transitions between two consecutive states of the lowest diagonal ladder can be driven by a $\sigma_+$ polarized RF field
\begin{align}
\label{Eq.: definition pulse}
\mathbf{F}_{\text{RF}}(t) = F_{\text{RF}}(t) \frac{\mathbf{e}_x + i \mathbf{e}_y}{2} +  F_{\text{RF}}^*(t) \frac{\mathbf{e}_x - i \mathbf{e}_y}{2},
\end{align}
with unitary vectors $\mathbf{e}_x$ and $\mathbf{e}_y$. The $x$- and $y$-components of $\mathbf{F}_{\text{RF}}$ correspond to the real and imaginary parts of the complex function $F_{\text{RF}}(t)$ and oscillate with a phase difference of $\pi/2$. The RF field couples to the atomic dipole moment $\hat{\mathbf{d}}$ via $\hat {h}_{\text{RF}}(t) = - \hat{\mathbf{d}} \cdot \mathbf{F}_{\text{RF}}(t)$. Unwanted $\sigma_-$ components may drive population out of the lowest diagonal ladder and reduce the circularization efficiency. To prevent this, we introduce a magnetic field $\text{B} = 14 \text{ G}$ along the $z$-direction. The Zeeman term $\hat{h}_{\text{B}}$ generates energy shifts proportional to $m$. Accordingly, the frequencies of the $\sigma_+$ (with $\Delta m =1$ and $ \Delta k =1$) and $\sigma_-$ (with $\Delta m =-1$ and $\Delta k =-1$) transitions are shifted in opposite directions. Then, if the RF field frequency is tuned in resonance with the $\sigma_+$ transitions, any accidental $\sigma_-$ polarization will not bring population out of the $\sigma_+$ lowest diagonal ladder as the $\sigma_-$ transitions are off-resonant. In total, the system is described by
\begin{align}
\label{Eq.: total single atom Hamiltonian}
    \hat{h}(t) = \hat{h}_0 + \hat{h}_{\mathcal{E}} + \hat{h}_{\text{B}} + \hat{h}_{\text{RF}}(t) \,,
\end{align}
where we make explicit that the only different ingredient when compared to the model considered in Refs. \cite{Larrouy2020,Patsch2018, Signoles2017} is the inclusion of the magnetic field. The static part $\hat{h}_\text{s} = \hat{h}_0 + \hat{h}_{\mathcal{E}} + \hat{h}_{\text{B}}$ gives rise to an harmonic energy spacing $\Delta E_{m \rightarrow m+1} = 3 n e a_0 \mathcal{E}/2 + \mu_B \text{B}$ in the lowest diagonal ladder \cite{Sakurai2017,Bethe1957, Kruckenhauser2022}, where $-e$ denotes the electron charge, $a_0$ the Bohr radius and $\mu_B$ the Bohr magneton. For our setting the transitions $m \rightarrow m+1$ in the lowest diagonal ladder correspond to an angular frequency of $\SI{229.2}{\mega\hertz}$. Besides, this harmonic ladder of states allows for the creation of minimum-uncertainty packets referred to as \textit{spin coherent states} \cite{Arecchi1972}. Both the $\ket{m=0}$ and the circular states belong to this class of coherent states, meaning that due to the closed non-dispersive evolution of coherent states under classical fields the angular momentum transfer during the circularization process can be seen as a rotation of states in a generalized Bloch sphere from minimum angular momentum projection ($\ket{m = 0}$, south pole) to the maxima ($\ket{52\,\text{C}}$, north pole) by means of a $\sigma_+$ polarized RF field $\pi$-pulse \cite{Facon2016,Larrouy2020,Patsch2018,Signoles2017,shore_book_1990, kam_2023,robert_2021}.

For rubidium most of the states for the valence electron follow a hydrogenic behavior as the respective orbitals are located far from the ionic core \cite{Kruckenhauser2022}. Only states with a low orbital angular momentum quantum number $l$ penetrate the core, such that shielding effects break down and the hydrogenic description is no longer valid. For states with $l < 4$, this results in a significant lowering of their corresponding eigenenergies, well described in the framework of \textit{quantum defects} \cite{gallagher_1994_book,Gallagher1988, Seaton1966}. In order to calculate the time evolution of the circularization process within the $n=52$ manifold of $^{87}$Rb, we construct $\hat{h}_\text{s}$ in the spherical basis with states $\ket{n,l,m}$, including quantum defects for states with $l\leq7$ and all states of neighbouring manifolds $n = 48, ..., 56$, and find its eigenstates through a numerical diagonalization. As shown in Fig.~\ref{fig_single_atom}(a) the energies of the states $\ket{m = 0}$ and  $\ket{m = 1}$ of $^{87}$Rb are significantly shifted when compared to the hydrogenic case. Only the transition from $\ket{m = 2}$ to $\ket{m = 3}$ is close to resonance with all the subsequent $m\to m+1$ transitions. This means that the subspace $\left\{\ket{m=2}, \ket{m=3}, ..., \ket{m=51} = \ket{52\text{C}}\right\}$ can be identified as the hydrogenic-like (almost harmonic) part of the lowest diagonal ladder allowing us to set $\ket{m=2}$ as the initial state for the circularization. In current circularization experiments, the $\ket{m=2}$ state is obtained through a two-photon laser excitation from the state $\ket{5S_{1/2}, F=2, m_F = 2}$, prepared by optical pumping in the presence of a magnetic field $B = 14 \, \text{G}$, to the state $\ket{52\,D_{5/2}, m_J = 5/2}$. The population is then transferred to $\ket{52\,F, m=2}$ with a microwave pulse, followed by adiabatically ramping up the electric field $\mathcal{E}$ from 0 to \SI{2.1}{\volt\per\centi\meter}~\cite{Ravon2023}.

Starting in the initial state $\ket{\psi(t=0)} = \ket{m = 2}$, we compute the time evolution $\ket{\psi(t)}$ by discretizing the RF pulse (setting it to constant values within time steps $\Delta t = \SI{0.1}{\nano\second}$) and solving the time-dependent Schrödinger equation for each time step subsequently. We consider pulses with sufficiently small components of $\sigma_-$ polarization so that we can effectively reduce the Hilbert space to the lowest five diagonal ladders during the complete time evolution, see appendices for further details. At the final time $t_f$, we obtain the evolved state $\ket{\psi(t_f)}$ that should be as close as possible to the circular state. Therefore, we search for pulse shapes $\textbf{F}_{\text{RF}}(t)$ that maximise the final circular state probability $p_{\text{C}} = \abs{\braket{n\,\text{C}}{\psi(t_f)}}^2$. From now on we denote pulses designed for single atoms as $\textbf{F}^\text{a}(t)$ and for atom pairs as $\textbf{F}^\text{aa}(t)$, where we omit the RF subscript to simplify notation. 

The state $\ket{m=2}$ chosen as initial state to avoid strong anharmonicities due to quantum defects \cite{Gallagher1988, Seaton1966} is not a spin coherent state. However, the $\pi$-pulse rotation relies on the closed evolution of coherent states under the action of a classical RF field. An optimal pulse therefore first transfers $\ket{m=2}$ to a nearby coherent state, which can be subsequently rotated similarly to the ideal hydrogenic case. This insight was the key factor in the optimization of pulses for an efficient circularization of a single rubidium atom carried out by S. Patsch and coauthors using quantum control methods \cite{Patsch2018}. The experimental implementation of the optimized pulses demonstrated a successful circularization by fast and coherent navigation of the $n=52$ Stark manifold of rubidium, achieving a fidelity of $\SI{96.2}{\percent}$ for total pulse durations of $\SI{125}{\nano\second}$ \cite{Larrouy2020}, thus reducing the circularization time by a factor of 30 with respect to the adiabatic passage. 

In what follows we adopt the approach presented in Ref. \cite{Patsch2018} to optimize RF pulses for the circularization in the absence of interactions. We additionally take into account a static magnetic field $B = 14 \, \text{G}$, which provides more robust results with respect to errors in the polarization of the RF field. For details on the numerical simulations see Appendix~\ref{Appendix: Details on simulation}. Based on the physical insight provided by the hydrogen-like model we choose as first initial guess pulse as a $\pi$-pulse with a flat-top shape with sine-square edges, a maximal amplitude of $\SI{90}{\milli\volt\per\centi\meter}$, and a duration of $t_\text{f} = \SI{65}{\nano\second}$. This duration arises from a balance between the duration of the pulse and the experimental feasibility of the amplitudes while keeping sufficient flexibility for the optimization. The circularization of rubidium starts in the state $\ket{m=2}$ and the transition towards $\ket{m=1}$ is not off-resonant enough to prevent a small population leakage which reduces the final circular state probability and that can be counteracted by lowering the amplitude at the beginning of the pulse \cite{Patsch2018}, see Appendix~\ref{Appendix: Guess Pulse}. The obtained single-atom initial guess pulse $\mathbf{F}_\text{guess}^\text{a}(t)$ leads to a final circular state probability $p_\text{C}$ of $\SI{90.1}{\percent}$. We rely on the vast toolbox offered by Quantum Optimal Control Theory (see Ref.~\cite{Koch2022} for a review) to perform a pulse optimization targeting to improve $p_{\text{C}}$ until it reaches a targeted fidelity of $p^\text{tg} = \SI{99}{\percent}$. Among a large set of pulse-shaping algorithms such as GRAPE \cite{Khaneja2005}, dCRAB \cite{Rach2015}, or Krotov's method \cite{Palao2003}, following S. Patsch and coauthors we use the latter since it is particularly suitable for the state preparation problem, see Appendix~\ref{Appendix: Krotov's method}. To ensure the feasibility of the obtained pulses we enforce experimental constraints during the optimization as explained in Appendix~\ref{Appendix: experimental constraints}; the experimental hardware limits maximal pulse amplitudes to $\SI{92}{\milli\volt\per\centi\meter}$ and the frequency bandwith to $\SI{480}{\mega\hertz}$ \cite{Patsch2018}. Moreover, to discuss the influence of the experimental constrains on the circularization, we examine pulses that fully comply with frequency and amplitude constraints (maximum amplitude of $\SI{92}{\milli\volt\per\centi\meter}$), pulses with enforced frequency constrains and relaxed amplitude constraints (maximum amplitudes of $\SI{130}{\milli\volt\per\centi\meter}$), and pulses with no constraints. 

Figure~\ref{fig_single_atom}(b) shows the pulse $\mathbf{F}_{\text{opt}}^{\text{a}}$ obtained after the optimization. The total pulse duration of $\SI{65}{\nano\second}$ implies a speed-up on the circularization process with respect to the adiabatic passage by a factor of approximately $60$ \cite{mehai2023}. By comparing the envelope of the guess pulse (shown in gray) to the optimized one ($x$- and $y$-components as solid and dashed lines) it is evident that the optimizer mainly alters the first $\SI{20}{\nano\second}$ of the pulse while leaving the rest mostly unchanged. This composition reflects the structure of the lowest diagonal ladder. For times below $\SI{20}{\nano\second}$, the population of the atomic state spreads in the low-$m$ states that are strongly affected by quantum defects and for which the pulse has to compensate for significant anharmonicities. Then, the population progressively \textit{climbs} the ladder following transitions that are harmonic up to first order Stark shifts as indicated by the increase in the expectation value of the projection of the angular momentum $\expval*{\hat{L}_z} \equiv \bar{m}(t)$ --black solid curve in Fig.~\ref{fig_single_atom}(c)-- from the lowest value $m=2$ to the highest one $m=51$. As explained in Refs.~\cite{Larrouy2020, Patsch2018}, the first stage of the pulse drives the state $\ket{m = 2}$ into a spin coherent state, which is subsequently rotated towards the circular state for times above $\SI{20}{\nano\second}$. Finally, the atom is prepared in a circular state with a probability of $p_{\text{C}} = \SI{99}{\percent}$. 


\section{A pair of interacting atoms}
\label{sec_atom_pair}
\begin{figure*}[t]
 \centering
 \includegraphics[width=\textwidth]{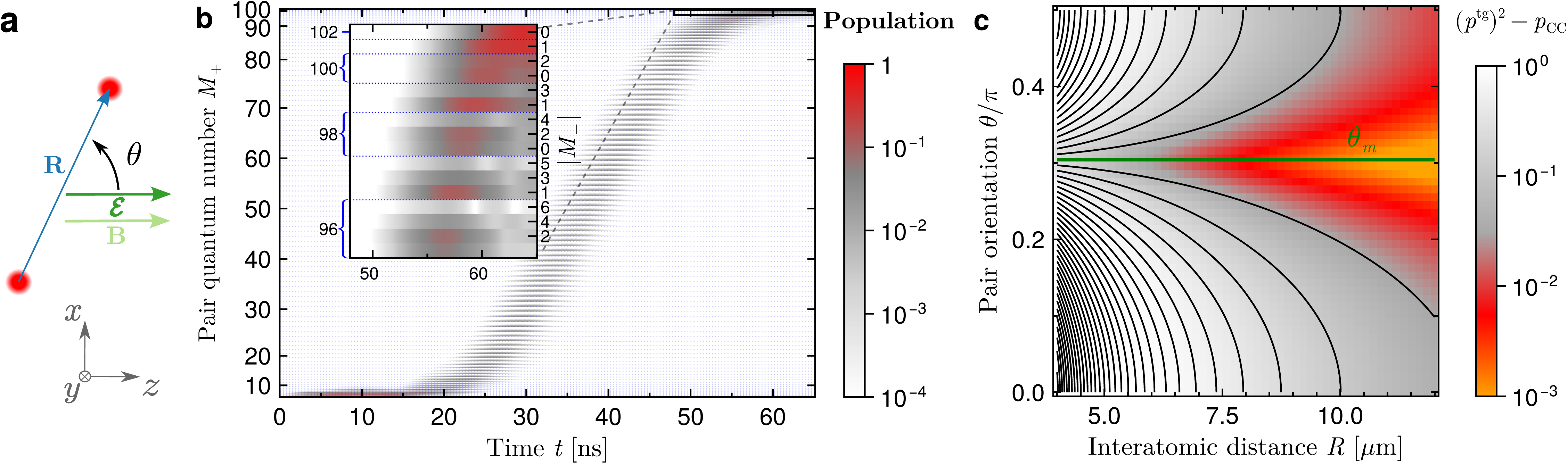}
   \caption{(a) A pair of Rydberg atoms with interatomic distance $R$ and angle $\theta$ with respect to the static electric and magnetic fields $\mathbfcal{E}$ and $\mathbf{B}$. (b) Angular momentum transfer for an interacting atom pair with $R = \SI{7}{\micro\meter}$ and $\theta = 0$ under the action of the RF pulse optimized for the circulrization of a single atom shown in Fig.~\ref{fig_single_atom}(b). The field magnitudes are set to $\mathcal{E}= \SI{2.1}{\volt\per\centi\meter}$ and $B = 14 \text{ G}$. The pair states are labeled by $M_+ = m_1 + m_2$ and $M_- = m_1 - m_2$ where $m_i$ with $i=1,2$ denote the single atom states belonging to the corresponding lowest diagonal ladder. During the evolution the population distribution is centered around pair states for which the difference of the magnetic quantum numbers $m_1$ and $m_2$ is small, allowing us to order them by increasing $M_-$ inside blocks (dotted blue lines) of fixed $M_+$. The interatomic interactions disturb the circularization such that the probability to reach the target state $\ket{n\text{C}; n\text{C}}$ is only $p_{\text{CC}} = \SI{32.3}{\percent}$. (c) Performance of the pulse optimized for the circularization of a single atom when applied to interacting pairs with different $R$ and $\theta$ quantified as the difference with respect to the highest possible final circular state probability of $\SI{98}{\percent}$. In general, the performance of the pulses follows the same scaling as the leading term of the interactions $(1-3 \cos^2\theta)/R^3$, depicted as black contour lines. As expected, the pulse optimized for the circularization of a single atom yields high $p_{\text{CC}}$ around the magic angle $\theta_{\text{m}} = \acos(1/\sqrt{3})$ and for large interatomic distances.}
 \label{fig_two_atoms_single_atom_pulse}
\end{figure*}
With the aim of optimizing the simultaneous circularization of interacting atoms, we consider two atoms separated by an interatomic distance $\mathbf{R}$ oriented with an angle $\theta$ with respect to the direction of the static electric and magnetic fields $\mathbfcal{E}$ and $\mathbf{B}$, see Fig.~\ref{fig_two_atoms_single_atom_pulse}(a). We are interested in the regime where $R$ ranges from 4 to $\SI{12}{\micro\meter}$, which leads to moderate interactions \cite{Nguyen2018, Cohen2021, mehai2023}. In this regime retardation effects can be neglected \cite{Deiglmayr2016}, and the atoms are distinguishable \cite{Weber2017, LeRoy1974, Ji1995}, such that we can model the interactions by the dipole-dipole term~\cite{Kruckenhauser2022, Deiglmayr2016,Weber2017, Flannery2005, Reinhard2007}
\begin{align}
    \label{eq: Interaction_hamiltonian}
    \hat{H}_{\text{int}} =\frac{1}{4 \pi \epsilon_0 R^3} \Big[&  \hat{\mathbf{d}}_1 \cdot \hat{\mathbf{d}}_1 - 3 \left(\hat{\mathbf{d}}_1 \cdot \mathbf{e}_R \right) \left( \hat{\mathbf{d}}_2 \cdot \mathbf{e}_R\right)\Big] \notag \\
    = \frac{e^2}{4 \pi \epsilon_0 R^3} \Big[&  \hat{x}_1 \hat{x}_2 (1 - 3 \sin^2\theta)+ \hat{z}_1 \hat{z}_2 (1 - 3 \cos^2\theta)  \notag \\
     & + \hat{y}_1 \hat{y}_2 - 3 \sin \theta \cos \theta (\hat{x}_1 \hat{z}_2 + \hat{z}_1 \hat{x}_2)\Big]
    ,
\end{align}
where $\epsilon_0$ is the vacuum permittivity and $\mathbf{e}_R = \mathbf{R}/R $. The sub-indices $i=1,2$ of the dipole operators $\hat{\mathbf{d}}_i$ and position operators $\hat{x}_i$, $\hat{y}_i$ and $\hat{z}_i$ indicate on which atom they act. We neglect the gradients of the static fields over the interatomic distance such that the fields couple to each atom individually. The atom pair system thus is described by the Hamiltonian
\begin{align}
\label{Eq.: total atom pair Hamiltonian}
    \hat{H}(t) & =  \hat{H}_\text{s} + \hat{H}_{\text{int}} + \hat{H}_{\text{RF}}(t) \notag \\
    & = \hat{H}_0 + \hat{H}_{\mathcal{E}} + \hat{H}_{\text{B}} + \hat{H}_{\text{int}} + \hat{H}_{\text{RF}}(t) \,,
\end{align}
where $\hat{H}_{\mathcal{E}} = \hat{h}_{{\mathcal{E}}, 1}\hat{\mathbb{I}}_2  + \hat{\mathbb{I}}_1 \hat{h}_{{\mathcal{E}}, 2}$ and similarly for $\hat{H}_{0}$, $\hat{H}_{B}$ and $\hat{H}_{\text{RF}}$. This Hamiltonian acts on \textit{pair states} of the form $\ket{\Psi} = \ket{\psi_1; \psi_2}$ and our goal is to find pulses $\mathbf{F}_{\text{opt}}^\text{aa}(t)$ that maximize the final circular pair state probability
\begin{align}
    p_{\text{CC}} = \abs{\braket{n\text{C}; n\text{C}}{\Psi(t_f)}}^2.
\end{align}
As for the single-atom case, we compute the evolution of the system under the action of the time dependent Hamiltonian of Eq.~\eqref{Eq.: total atom pair Hamiltonian}, assuming both atoms initially prepared in the lowest diagonal ladder state $\ket{m = 2}$, i.e., $\ket{\Psi(t=0)} = \ket{m_1 = 2; m_2 = 2}$, see Appendices \ref{Appendix:couplings induced by interactions} and \ref{Appendix: state space reduction} for details on the time evolution simulations. Here and in the following, we denote pair states $\ket{m_1;m_2}$, where both atoms $i=1,2$ populate a lowest diagonal ladder state $\ket{m_i}$, by  $\ket{M_+, M_-}$, where $M_{\pm} = m_1 \pm m_2$ quantify the sum and difference of the respective magnetic quantum numbers $m_i$.

To illustrate how the interatomic interactions interfere with the circularization we evolve an atom pair with the pulse $\mathbf{F}_{\text{opt}}^\text{a}(t)$ optimized for the circularization of a single atom. The highest possible final circular pair state probability that can be reached by this pulse is $p_\text{CC} = (p^\text{tg})^2= \SI{98}{\percent}$, corresponding to a non-interacting atom pair. In the following, we focus on an atom pair with $R = \SI{7}{\micro\meter}$ and $\theta = 0$. The resulting time evolution is shown in Figure~\ref{fig_two_atoms_single_atom_pulse}(b). The atoms populate pair states with an increasing total magnetic quantum number $M_+$, gradually evolving towards the circular pair state $\ket{M_+ = 2(n-1), M_- = 0}$. However, as highlighted in the inset of Fig.~\ref{fig_two_atoms_single_atom_pulse}(b), the interactions interfere with the circularization. A large part of the population remains in high but not maximum $M_+$ states, leading to a reduction of the final circular pair state probability to $p_\text{CC} = \SI{32.3}{\percent}$. Figure~\ref{fig_two_atoms_single_atom_pulse}(c) depicts the difference between the highest possible final circular pair state probability of $\SI{98}{\percent}$ and the obtained probability as a function of the interatomic distance $R$ and the angle $\theta$ between the interatomic axis and the static fields. Pairs arranged at the magic angle, $\theta_\text{m} = \arccos{(1/\sqrt{3})}$, reach circular pair state probabilities $p_{\text{CC}} \geq \SI{97}{\percent}$ if they are separated by $R \geq \SI{7.2}{\micro\meter}$. The same pulse acting on pairs separated by $\SI{10}{\micro\meter}$ with $\theta = 0$ yields $p_{\text{CC}} \geq \SI{85.9}{\percent}$. The highest possible performance of $\SI{98}{\percent}$ is reached for any angle $\theta$ for interatomic distances larger than $\SI{12.5}{\micro\meter}$ -- not shown in Fig.~\ref{fig_two_atoms_single_atom_pulse}(c). The value of $p_{\text{CC}}$ follows a $(1-3\cos^2\theta)/R^3$ scaling (black contour lines) given by the coupling strength between angular-momentum-preserving terms of the dipole-dipole interactions. 

\begin{figure*}[t]
 \centering
 \includegraphics[width=\textwidth]{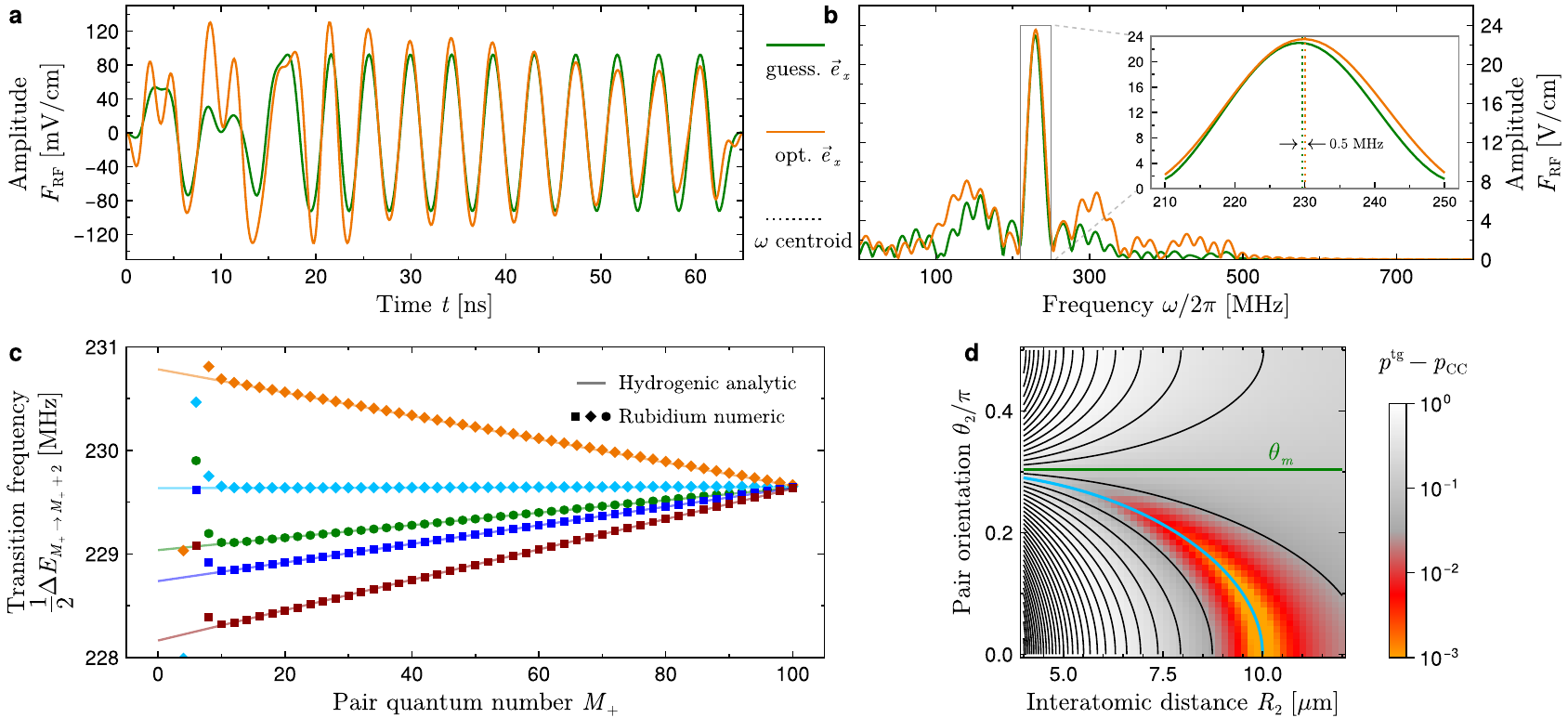}
   \caption{Main mechanism through which the interactions disturb the circularization. (a-b) Comparison between the RF pulse optimized for the circularization of a single atom (green) with a pulse optimized for the simultaneous circularization of a pair of atoms (orange) with $R = \SI{7}{\micro\meter}$ and $\theta = 0$, targeting in both cases to an efficiency $p^\text{tg}=99\%$. The static electric and magnetic field have magnitudes $\mathcal{E}= \SI{2.1}{\volt\per\centi\meter}$ and $B = 14 \text{ G}$. Panel (a) depicts the $x$-component of both pulses ($y$-components exhibit similar behavior) while panel (b) shows the corresponding spectral analysis. The Fourier transform of the pulse optimized for the circularization of a single atom exhibits a dominant peak around the mean lowest diagonal ladder transition frequency, which is shifted by $\SI{0.5}{\mega\hertz}$ towards higher frequencies in the pulse optimized for the circularization of the interacting pair, see inset. (c) The numerical calculation of the transition frequencies of Rubidium (points) closely matches the linear hydrogenic analytical expression of Eq.~\eqref{eq: lowest diagonal pair ladder frequencies} (solid lines) for $M_+ \geq 10$. At the magic angle $\theta_\text{m}$ the ladder is equivalent to the non-interacting case obtained for sufficiently large $R$. (d) Performance of a pulse optimized using Krotov's method for an atom pair with interatomic distance $R_1=\SI{10}{\micro\meter}$ and angle $\theta_1=0$ when applied to a general arrangement $(R_2, \theta_2)$. The performance of the pulse follows the same scaling as the leading term of the interactions $(1-3 \cos^2\theta)/R^3$ (black contour lines) and is retained for all the arrangements fulfilling Eq.~\eqref{eq: condition similar lowest diagonal pair ladder} (blue curves). All the presented pulses have the same duration of $\SI{65}{\nano\second}$.}
 \label{fig_optimized_pulse_transitions}
\end{figure*}

As a benchmark, we first employ Krotov’s method with $\mathbf{F}_{\text{opt}}^{\text{a}}$ as initial guess to obtain pulses targeting to the maximization of the final circular pair state probability for an atom pair with $R = \SI{7}{\micro\meter}$ and $\theta = 0$, see Fig.~\ref{fig_optimized_pulse_transitions}. When enforcing the experimental maximum amplitude constraint of $\SI{92}{\milli\volt\per\centi\meter}$, the optimizer converges to $p_{\text{CC}}=\SI{96.8}{\percent}$. This can be increased to our target of $\SI{99}{\percent}$ by allowing for longer pulse durations or by relaxing the experimental amplitude constraint. For example, we reach $p_\text{CC} = \SI{99}{\percent}$ when allowing maximum amplitudes of $\SI{130}{\milli\volt\per\centi\meter}$, as shown in Fig.~\ref{fig_optimized_pulse_transitions}(a). To assess the main differences between $\mathbf{F}_{\text{opt}}^{\text{aa}}$ and $\mathbf{F}_{\text{opt}}^{\text{a}}$ we perform their spectral analysis. The frequency spectrum of $\mathbf{F}_{\text{opt}}^{\text{a}}$ reveals a peak around the mean lowest-diagonal-ladder transition frequency, shifted to higher frequencies in the case of $\mathbf{F}_{\text{opt}}^{\text{aa}}$, see Fig.~\ref{fig_optimized_pulse_transitions}(b). This frequency shift increases with interaction strength and changes its sign for a pair arrangement with $\theta = \pi/2$, corresponding to the attractive or repulsive interactions of a head-to-tail or side-by-side arrangement of two dipoles. 

\newpage
\section{Frequency shifts induced by the interactions}
\label{sec_frequency_shifts}

In this section we draw on the hydrogenic approximation to develop an understanding for the origin of the observed frequency shifts. In the absence of interactions, the two independent atoms simultaneously climb their respective lowest diagonal ladder from $\ket{m = 2}$ towards the circular state following the evolution obtained for a single atom, i.e., $\bar{m}_1 (t) = \bar{m}_2 (t) = \bar{m} (t)$ for all $t$, see Fig.~\ref{fig_single_atom}(c). In the presence of dipole-dipole interactions the total Hamiltonian is symmetric under the exchange of particles and the evolution preserves the symmetry of the states. Since our considered evolution starts in a symmetric states, the pair state remains symmetric for all times implying that $\expval*{\hat{L}^{z}_{1}-\hat{L}^{z}_2} = 0$, or equivalently $\bar{m}_1(t) = \bar{m}_2 (t)$. We then focus on the pair states for which $M_+ = 2m$ and $M_- = 0$, denoted as $\ket{M_+}$ in the following, and calculate the interaction-induced energy shifts for the transitions with $\Delta M_+ = 2$ and $\Delta M_- = 0$, i.e. considering only two simultaneous single atom transitions with $\Delta m = 1$. The energy of the pair states $\ket{M_+}$ is given by the expectation value of the operator $\hat{H}_0 + \hat{H}_{\mathcal{E}} + \hat{H}_{\text{B}} +  e^2(1 - 3 \cos^2\theta)\hat{z}_1 \hat{z}_2 / (4 \pi \epsilon_0 R^3)$ and the energy associated to the $\Delta M_+ = 2$ and $\Delta M_- = 0$ transition reads
\begin{align}
\label{eq: lowest diagonal pair ladder frequencies}
    \Delta E_{M_+ \rightarrow M_+ + 2} 
    =& \langle  M_+ + 2 | \hat{H}_{\text{eff}} | M_+ + 2 \rangle  -  \langle M_+  | \hat{H}_{\text{eff}} | M_
    +\rangle  \notag \\
    =& \ 3 e a_0 n \mathcal{E} +2\mu_B B \\ 
    & + 3 \pi \epsilon_0 a_0^3 n^4 \mathcal{E}^2\left( - n + 2M_+ + 3\right)  \notag \\ 
    & + \frac{9 e^2 a_0^2 n^2}{16 \pi \varepsilon_0} \frac{1-3\cos^2\theta}{R^3} \left( -2n + M_++ 3\right). \notag
\end{align}
Due to the second-order Stark shift and the dipole-dipole interactions we obtain a linear function of $M_+$ parametrized by
\begin{align}
\label{eq_ldl_pair_transitions_coefs}
    \alpha_0 & = \frac{3}{2} e a_0 n \mathcal{E} + \mu_B B + \frac{3}{2} \pi  \epsilon_0 a_0^3 n^4 \mathcal{E}^2\left( - n + 3\right)\,, \notag \\
    \beta_0 & = 3 \pi \epsilon_0 a_0^3 n^4\mathcal{E}^2 \,, \notag \\
    \alpha_\text{int} & = \frac{9 e^2 a_0^2 n^2}{32 \pi \epsilon_0} \frac{1-3\cos^2\theta}{R^3} \left( -2n + 3\right) \,,\notag \\
    \beta_\text{int} & = \frac{9 e^2 a_0^2 n^2}{32 \pi \epsilon_0} \frac{1-3\cos^2\theta}{R^3}  \, , 
\end{align}
such that $\alpha_\text{int}+ \beta_\text{int} M_+$ gives the energy shift per atom induced by the interactions. In other words, we can associate to each pair state $\ket{M_+}$ a shift in the transition frequency due to the dipole-dipole interaction given by $\Delta \omega (M_+) = \left(\alpha_\text{int} + \beta_\text{int} M_+ \right) / \hbar$, where $\hbar$ is the reduced Planck constant. These shifts bring the optimized single-atom pulse out of resonance thus reducing the final circular pair state probability. As mentioned before, the optimization with Krotov's method shifts the leading frequency of $\mathbf{F}_{\text{opt}}^{\text{a}}$. This shift increases for stronger interactions (smaller interatomic distances) and changes its sign according to $(1-3\cos^2\theta)$ vanishing for $\theta = \theta_\text{m}$. The behavior of the frequency shift with respect to the parameters of the interactions is in agreement with that of the corresponding transition energies numerically obtained for Rubidium (see Fig.~\ref{fig_optimized_pulse_transitions}(c)) which, in turn, match closely the analytical hydrogenic expression of Eq.~\eqref{eq: lowest diagonal pair ladder frequencies} with the exception of the low-$M_+$ states affected by quantum defects. To sum up, Eq.~\eqref{eq: lowest diagonal pair ladder frequencies} captures the main effect of the interactions as a shift in the relevant two-body transitions that the optimizer counteracts by introducing a similar shift in the leading frequency of the pulse.

Equation~\eqref{eq: lowest diagonal pair ladder frequencies} also provides a way to identify atom pair arrangements that lead to similar shifts in the transition energies. Two different arrangements $(R_1, \theta_1)$ and $(R_2, \theta_2)$ lead to the same shifts as long as
\begin{equation}
\label{eq: condition similar lowest diagonal pair ladder}
   \frac{1 - 3 \cos^2\theta_1}{R_1^3} = \frac{1 - 3 \cos^2\theta_2}{R_2^3} \, ,
\end{equation}
therefore, a pulse optimized for $(R_1, \theta_1)$ is expected to retain its performance for all the configurations $(R_2, \theta_2)$ satisfying the latter equation because it compensates for similar shifts. This is shown in Fig.~\ref{fig_optimized_pulse_transitions}(d) for a pulse optimized with Krotov's method under relaxed amplitude constraints up to $p_{\text{CC}} = 99\%$ for $(R_1 = \SI{10}{\micro\meter}, \theta_1 = 0$). The same pulse reaches a similar performance for other configurations $(R_2, \theta_2)$ that fulfill Eq.~\eqref{eq: condition similar lowest diagonal pair ladder}, depicted as a light blue curve. Only when $R$ becomes very small its performance degrades, yielding $p_{\text{CC}} \leq \SI{90}{\percent}$ for $R_2 \leq \SI{5}{\micro\meter}$. This is due to the neglected contributions of all the transitions with $\Delta M_+ \neq 2$ and $\Delta M_- \neq 0$.

Based on the $R^{-3}$ scaling of the interactions we expect the error associated to a fixed relative variation of $R$ to increase for shorter interatomic distances. For example, the pulse optimized for ($R_1 = \SI{10}{\micro\meter}, \ \theta_1 = 0$) leads to a difference in $p_{\text{CC}}$ of $\SI{0.9}{\percent}$ when applied to configurations ($R_2 = R_1 + \SI{1}{\micro\meter}, \ \theta_2 = \theta_1 + 0.05 \cdot \pi$). In contrast, this difference increases to $\SI{4.9}{\percent}$ when using a pulse optimized for $(R_1 = \SI{7}{\micro\meter},\ \theta_1 =0$) for arrangements ($R_2 = R_1 + \SI{0.7}{\micro\meter}, \ \theta_2 = \theta_1 + 0.05 \cdot \pi$).

\section{The adaptive frequency approach}
\label{sec_ad_freq}

\begin{figure*}[t]
 \centering
 \includegraphics[width=\textwidth]{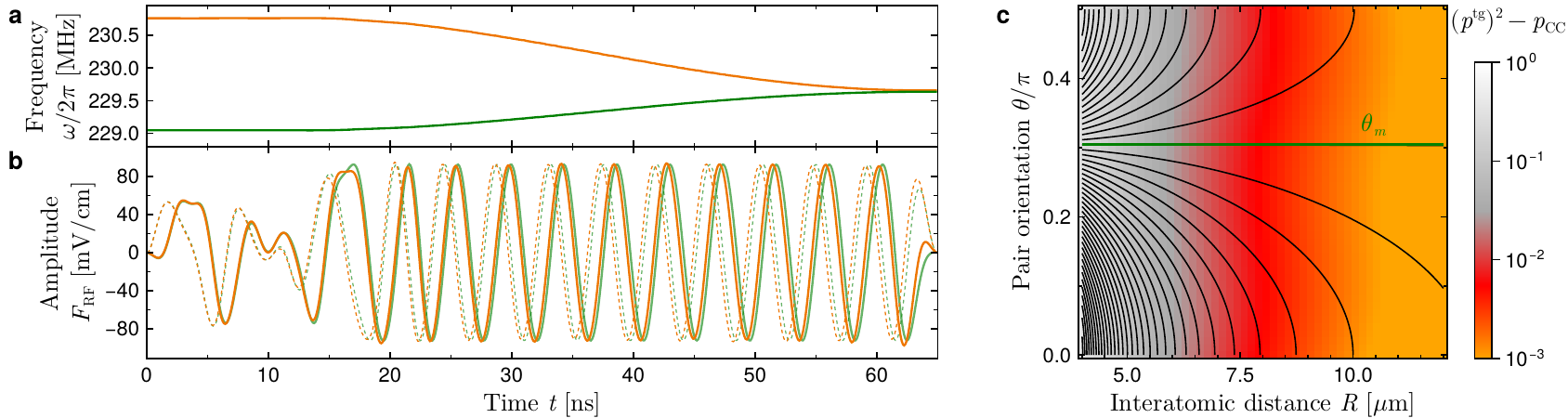}
   \caption{Analytical adaptation of the single-atom pulse to two interacting atoms. (a) Frequencies associated to the transitions supported by the lowest diagonal ladders for two non-interacting (green) and interacting (orange) hydrogenic atoms. (b) $x-$ and $y-$ components (solid and dashed orange curves, respectively) of the analytical adapted pulse $\mathbf{F}_{\text{ad. freq.}}^\text{aa}$ for interacting atoms obtained by incorporating into $\mathbf{F}_{\text{Krotov}}^\text{a}$ (in green) a phase shift that counteracts the main effect of interactions, see Eq.~\eqref{eq: adapted frequency pulse}. The resulting adapted pulse leads to $p_{\text{CC}} = \SI{96.9}{\percent}$ for $R = \SI{7}{\micro\meter}$ and $\theta = 0$. (c) Differences between the maximum achievable final circular pair probability of $(p^\text{tg})^2 = 98\%$ and the one obtained after evolving the system with the adapted pulse for each $R$ and $\theta$. The black contour lines indicate the contour curves for $(1-3 \cos^2\theta)/R^3$ corresponding to the leading term of the interactions. All the depicted pulses have the same duration of $\SI{65}{\nano\second}$.}
 \label{fig_adapted_pulse}
\end{figure*}

The obtained frequency shifts allow us to propose an adaptation of the single-atom pulse to the interacting case. Our proposal incorporates a phase modulation that accounts for the frequency shifts induced by the interactions
\begin{align}
\label{eq: adapted frequency pulse}
   \text{F}_{\text{ad. freq.}}^\text{aa}(t) 
   & = \text{F}_{\text{opt}}^\text{a}(t) \, e^{-i \int_0^t\Delta \omega(\tau) \text{d}\tau}  \\
   & = \text{F}_{\text{opt}}^\text{a}(t) \, e^{-i \left( \alpha_\text{int} \, t + \beta_\text{int} \int_0^t  2 (\bar{m}_\text{opt}(\tau)-1) \text{d}\tau \right)/\hbar } \,, \notag
\end{align}
where $\bar{m}_\text{opt}(t)$ denotes the evolution of the expectation value of the angular momentum projection of a single atom driven by the optimized single-atom pulse $\mathbf{F}_{\text{opt}}^\text{a}(t)$ of Fig.~\ref{fig_single_atom}(b), and the unitary offset is introduced to satisfy $\Delta\omega (t_\text{f}) =0$, see Figure~\ref{fig_adapted_pulse}(a). Note that since the single-atom pulse is tailored to take the quantum defects into account and satisfies the amplitude constraints, the adapted pulse carries over both features by construction. Furthermore, because $\alpha_\text{int}$ and $\beta_\text{int}$ vanish for large interatomic distances or for $\theta = \theta_m$, the adapted pulse reduces to the pulse optimized for a single particle in the non-interacting limit.

Figure~\ref{fig_adapted_pulse}(b) depicts the adapted pulse $\mathbf{F}_{\text{ad. freq.}}^\text{aa}$ obtained for the atom pair arranged at $(R = \SI{7}{\micro\meter}, \ \theta = 0)$. This pulse leads to a final circular pair state probability of $\SI{96.9}{\percent}$, which is comparable to the $p_{\text{CC}}=\SI{96.8}{\percent}$ obtained via an optimization with Krotov's method using experimental constraints and the same single-atom pulse as initial guess. We further construct adapted pulses for pairs with different $R$ and $\theta$ and show the obtained $p_{\text{CC}}$ in Fig.~\ref{fig_adapted_pulse}(c). Given the simplicity of the proposed correction, its performance is remarkable; our adapted pulse provides for a final circular pair probability higher than $\SI{95}{\percent}$ ($\SI{97}{\percent}$) for interatomic distances larger than $\SI{6.5}{\micro\meter}$ ($\SI{7.6}{\micro\meter}$) and any angular orientation. 

The analytical derivation of the linear parameters given in Eq.~\eqref{eq_ldl_pair_transitions_coefs} rely both on the hydrogenic approximation and on the physical picture of two atoms simultaneously climbing their respective lowest diagonal ladder while populating only pair states with $m_1=m_2$ which is equivalent to considering only two atom transitions composed of two simultaneous single atom transitions with $m \to m+1$. However, the latter approximation breaks down for strong interactions. To overcome this we also perform an optimization\footnote{We apply the algorithm BFGS \cite{BROYDEN1970, Fletcher1970, Goldfarb1970} as implemented in the software package \texttt{Optim.jl} \cite{Optim}.} of $\alpha_\text{int}$ and $\beta_\text{int}$ using as initial guess the analytical parameters provided by Eq.~\eqref{eq_ldl_pair_transitions_coefs}. When compared to the adapted pulse constructed with the analytical $\alpha_\text{int}$ and $\beta_\text{int}$, the pulses resulting from the optimization of these linear parameters lead to a marginal improvement below $\SI{0.1}{\percent}$ for moderate interactions $(R = \SI{7}{\micro\meter}, \ \theta = 0)$.  

\section{The optimization scheme}
\label{sec_optimization}

\begin{figure*}[t]
 \centering
 \includegraphics[width=\textwidth]{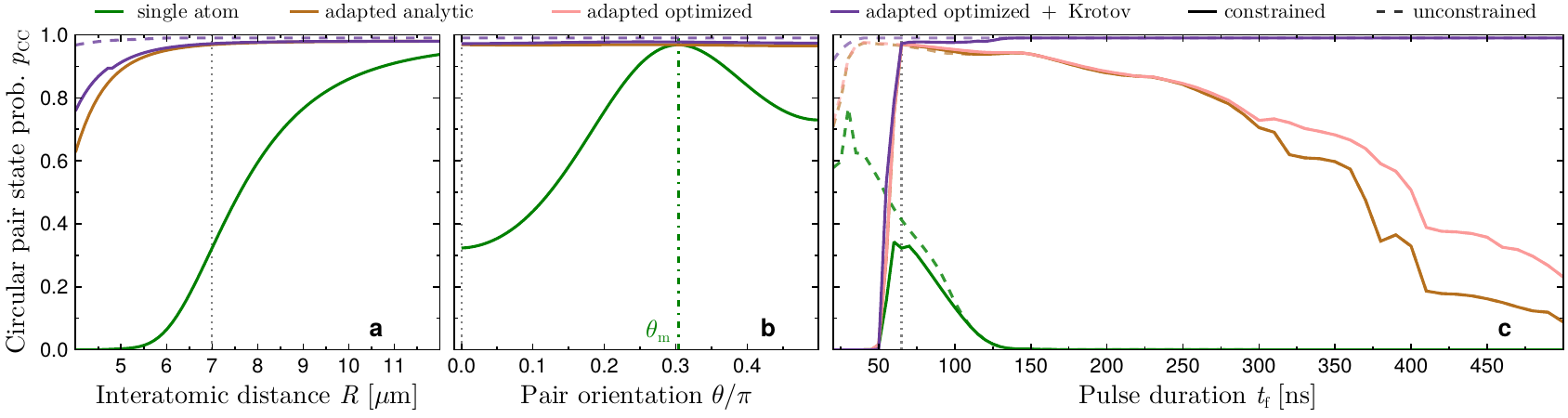}
   \caption{Final circular pair state probability $p_\text{CC}$ obtained after the evolution of an atom pair under the action of different pulses (a) as a function of the interatomic distancde $R$ for a fixed angle $\theta = 0$ and pulse duration $t_{\text{f}} = \SI{65}{\nano \second}$, (b) as a function of $\theta$ for a fixed $R = \SI{7}{\micro\meter}$ and $t_{\text{f}} = \SI{65}{\nano \second}$, and (c) as a function of $t_{\text{f}}$ for fixed $R = \SI{7}{\micro\meter}$ and $\theta = 0$. The dotted vertical lines in each panel indicate the value to which the corresponding parameter is fixed in the remaining panels. As expected, the optimized single-atom pulse of Fig.~\ref{fig_single_atom}(b) (green) results in high values of $p_{\text{CC}}$ around $\theta_{\text{m}}$ and for large $R$. The optimized single-atom pulse is then adapted to the interactions as indicated by Eq.~\eqref{eq: adapted frequency pulse} using either the analytical hydrogenic parameters of Eq.~\eqref{eq_ldl_pair_transitions_coefs} (brown) or optimized linear parameters $\alpha_\text{int}$ and $\beta_\text{int}$ (salmon pink). The adapted pulses with optimized parameters are further optimized using Krotov's method (purple). Results are shown for pulses with and without enforced experimental constrains (solid and dashed). The adapted pulses allow for an improvement in $p_\text{CC}$ with respect to the single atom pulse of at least $\SI{40}{\percent}$ for $R <\SI{8}{\micro\meter}$ and for any angle $\theta$. For $R=\SI{7}{\micro\meter}$ and $\theta=0$ our adapted pulse constructed with analytical parameters leads to $p_{\text{CC}} > \SI{90}{\percent}$ for $t_{\text{f}} < \SI{200}{\nano\second}$. The adapted pulse with optimized parameters outperforms the adapted pulse with analytical parameters for $t_{\text{f}} >\SI{300}{\nano\second}$. Using Krotov's method to further shape our adapted pulses results in high-fidelity pulses for all $t_{\text{f}} >\SI{60}{\nano\second}$ with enforced experimental constraints and above $\SI{40}{\nano\second}$ without them.}
 \label{fig_comparison_approaches}
\end{figure*}

In the previous section we presented an approach to adapt single-particle pulses to interacting atoms. We showed that the obtained analytical parametrization of the adapted pulse leads to high final circular pair state probabilities by itself and also provides a way to shape the pulses by optimizing over only two parameters. In what follows, we compare the performance of the adapted pulse approach with and without a subsequent an optimization via Krotov's method. Based on this comparison we propose a workflow to speed up the optimization of pulses for the circularization of interacting atoms by identifying regimes for which the straightforward use of the adapted pulse allows to avoid the computationally expensive exhaustive search over a large range of possible pulses made by Krotov's method. When the fidelity of the adapted pulse is less than the target, it still provides for a good initial guess, saving significant computational time in a second (or third, if the linear parameters of the adapted pulse are also optimized) optimization layer using Krotov's method.

Figure~\ref{fig_comparison_approaches} shows the final circular pair state probability obtained after the evolution of the interacting atom pair under the action of various pulses. We compare the performance of pulses optimized for the circulatization of a single atom (green curves), adapted pulses with analytical (brown) and optimized (salmon pink) parameters, and pulses obtained via Krotov's method (purple) using the adapted pulse with optimized parameters as initial guess. Alongside pulses that comply with the experimental constraints (solid), we also show pulses obtained without enforcing any constraint (dashed). The results are shown as a function of $R$ with $\theta = 0$ in panel (a), as a function of $\theta$ for fixed $R = \SI{7}{\micro\meter}$ in panel (b), and as a function of the total pulse duration $t_\text{f}$ between 25 and $\SI{500}{\nano\second}$ for $R = \SI{7}{\micro\meter}$ and $\theta = 0$ in panel (c). 

As expected, the single-atom pulse $\mathbf{F}_{\text{opt}}^\text{a}$ performs well in the regime of weak interactions. For a distance of $\SI{12}{\micro\meter}$ it yields $p_{\text{CC}} = \SI{93.7}{\percent}$, which deteriorates for smaller interatomic distances until vanishing for $R \leq \SI{4.5}{\micro\meter}$, see Fig.~\ref{fig_comparison_approaches}(a). The detrimental effects of the interactions on the circularization process can be counteracted by using adapted pulses constructed with the hydrogenic analytical parameters $\alpha_\text{int}$ and $\beta_\text{int}$. These pulses improve upon $\mathbf{F}_{\text{opt}}^\text{a}$ by at least $\SI{70}{\percent}$ for $R = \SI{5}{\micro\meter}$ and reach the highest possible final circular probability of $98\%$ for $R \geq \SI{12}{\micro\meter}$. The performance of the adapted pulses deteriorates for $R < \SI{6.5}{\micro\meter}$, yielding $p_\text{CC} = \SI{62.7}{\percent}$ for $R = \SI{4}{\micro\meter}$. This deterioration is due to the contribution of the transitions not considered in our simplified picture of a synchronous evolution populating only states with $m_1=m_2$ that reduces the two atom transitions to the set of two simultaneous single atom transitions with $m \to m+1$. An optimization of the parameters $\alpha_\text{int}$ and $\beta_\text{int}$ leads to an improvement below $\SI{0.1}{\percent}$ for all $R$, suggesting that the analytical parameters are close to optimal. A further optimization with Krotov's method considering experimental amplitude constraints and using the adapted pulse with optimized parameters as initial guess leads to a difference in $p_{\text{CC}}$ below $\SI{1}{\percent}$ for $R >\SI{5.7}{\micro\meter}$. The same difference is below $\SI{0.1}{\percent}$ for $R > \SI{8.2}{\micro\meter}$. When using Krotov's method without constraints the target fidelity of $99\%$ is reached for $R > \SI{5.9}{\micro\meter}$.

As shown in Fig.~\ref{fig_comparison_approaches}(b), when fixing $R = \SI{7}{\micro\meter}$ and varying $\theta$ from $0$ to $\pi/2$ the single-atom pulse $\mathbf{F}_{\text{opt}}^\text{a}$ reaches high values of $p_{\text{CC}}$ around the magic angle, yielding $p_\text{CC} = \SI{96.8}{\percent}$ for $\theta = \theta_\text{m}$. This performance decays towards $\theta =0$. The difference in $p_{\text{CC}}$ obtained with the adapted analytical pulse or with optimized parameters remains below $\SI{0.1}{\percent}$ for any value of $\theta$. A further optimization of the adapted pulse with optimized parameters via Krotov's method leads to improvements below $\SI{1}{\percent}$ if experimental amplitude constraints are enforced, but reaches the target fidelity of $99\%$ for any angle $\theta$ without constraints. 
\begin{figure*}[t] 
    \centering
\includegraphics[width=1\textwidth]{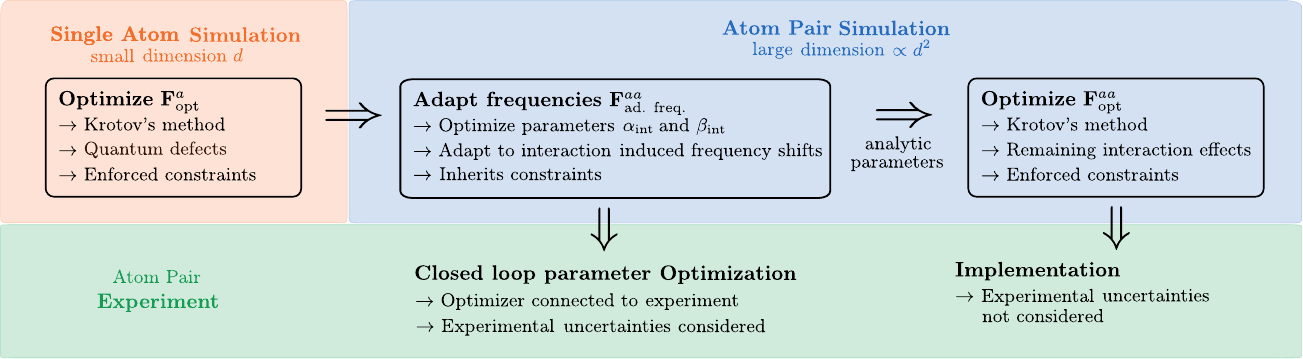} 
    \caption{RF-pulse optimization strategy for circularizing two interacting Rubidium atoms. A pulse is first optimized in a single-atom simulation to consider quantum defects, then adapted to interaction-induced frequency shifts using the analytical parameters $\alpha_{\mathrm{int}}$ and $\beta_{\mathrm{int}}$. The resulting pulse can be further refined either in a closed loop directly connected to the experiment or through two-atom simulations using pulse shaping algorithms such as Krotov's method.}
    \label{fig: optimization_scheme} 
\end{figure*}

Since our main goal is to find fast and accurate pulses that satisfy the experimental constraints, we analyze the influence of the pulse duration $t_{\text{f}}$ on the circularization process, see Fig.~\ref{fig_comparison_approaches}(c). For increasing duration of the pulse the spectral peak narrows, such that the frequency shifts induced by the interactions become increasingly relevant thus deteriorating the circularization process. The latter results in a decreasing performance of $\mathbf{F}_{\text{opt}}^\text{a}$ for long durations. In particular, $p_{\text{CC}}$ vanishes for $t_{\text{f}} > \SI{180}{\nano\second}$. In contrast, the adapted pulse constructed with the analytical linear parameters yields $p_{\text{CC}} >\SI{90}{\percent}$ for $t_{\text{f}} <\SI{200}{\nano\second}$, which constitutes a remarkable achievement given the simplicity of our proposal. Adapted pulses based on analytical parameters are outperformed by those constructed with the optimized linear parameters $\alpha_\text{int}$ and $\beta_\text{int}$ for $t_{\text{f}} > \SI{290}{\nano\second}$. As mentioned above, for $t_{\text{f}} > \SI{180}{\nano\second}$ the performance of $\mathbf{F}_{\text{opt}}^\text{a}$  deteriorates to such an extent that it is no longer a suitable initial guess pulse for a subsequent optimization with Krotov's method. To overcome extremely slow convergences, we use the adapted pulse with optimized parameters as initial guess for Krotov's algorithm and recover high fidelity pulses whose performance is only limited by experimental constraints. The obtained pulses yield $p_{\text{CC}} >\SI{96.9}{\percent}$ for $t_{\text{f}} > \SI{65}{\nano\second}$ if amplitude constraints are enforced. Moreover, for $t_{\text{f}} > \SI{135}{\nano\second}$ we reach our target fidelity of $99\%$. The performance of the pulses shaped via Krotov's method decays for $t_{\text{f}} <\SI{60}{\nano\second}$, as they are too short to transfer all the population to the circular pair state. When lifting the constraints during Krotov's optimizations we are able to find pulses yielding $p_\text{CC} =\SI{99}{\percent}$ for $t_{\text{f}} > \SI{40}{\nano\second}$. Overall, the experimental constraints imposed during pulse-shaping are the limiting factor for achieving high-fidelity circular pair states for short pulse durations or strong interactions. When constraints are not enforced, optimizations based on Krotov's method using the adapted pulses as initial guess reach a fidelity of $99\%$ by inducing maximum amplitudes that can exceed our chosen experimental maximum by a factor of five, see Appendix~\ref{Appendix: experimental constraints}. 

Based on all of the above we propose an optimization strategy that is illustraded schematically in Fig.~\ref{fig: optimization_scheme}. First, optimize the single-atom pulse taking into account all the experimental constraints and the quantum defects. Then, construct the adapted pulse using Eq.~\eqref{eq: adapted frequency pulse} and the analytical hydrogenic parameters of Eq.~\eqref{eq_ldl_pair_transitions_coefs}. The latter step leads to final circular pair state populations higher than $\SI{95}{\percent}$ for atomic arrangements with interatomic distances larger than $\SI{6.5}{\micro\meter}$ and pulse durations below $\SI{100}{\nano\second}$ without additional computational cost. If a higher circular pair state probability is desired, the next step with minimal computational cost is to optimize the linear parameters $\alpha_\text{int}$ and $\beta_\text{int}$. The last optimization layer uses the pulse obtained in the previous step as initial guess for Krotov's method. This final option carries the highest computational cost, since it relies on simulating a large state space and performing an exhaustive search over pulse shapes. Overall, for short pulse durations the adapted pulse with the analytical parameters results in high fidelity pulses without the need of extra optimizations. However, for interatomic distances below $\SI{6.5}{\micro\meter}$ or when the pulse duration increases, additional optimizations steps allow for further improvements.

\section{Summary and Conclusions}
\label{sec_conclus}

In this work, we address the challenge of fast and precise preparation of circular states in interacting Rydberg atoms. The last stage of this preparation, also known as \textit{circularization}, involves the transfer of a large amount of angular momentum quanta to the valence electron of the Rydberg atom and has been identified as a major bottleneck for all the technological applications of circular states, ranging from quantum simulation and computation \cite{Nguyen2018,Cohen2021}, to quantum sensing \cite{Dietsche2019}. 

In current experiments, the circularization is typically implemented via an adiabatic passage, yielding circular state probabilities of up to \SI{99.5}{\percent} for non-interacting atoms. As a faster alternative, short RF pulses with durations of around \SI{100}{\nano\second} have been developed for non-interacting atoms and successfully applied to dilute atomic clouds, where interactions are negligible, achieving fidelities of \SI{96}{\percent} \cite{Larrouy2020}. Both approaches therefore avoid one of the main ingredients required for quantum simulation and quantum computation: the interactions. With this work, we extend the design of fast RF pulses to interacting atoms and propose a method to adapt the single-atom pulses developed in Ref.~\cite{Patsch2018} by incorporating the leading effects of interactions. By combining quantum optimal control techniques with the analytical derivation of a pulse correction due to interactions, we obtain high-fidelity pulses allowing for a fast circularization, while at the same time complying with experimental bandwidth and maximum amplitude constraints. Compared to the currently used adiabatic passage technique, the optimized pulses speed-up the circularization task by a factor of $60$ \cite{mehai2023}.

Our proposal relies on a intuitive physical picture. Based on conserved quantities we identify the relevant transitions for two interacting atoms as those composed by two simultaneous and equivalent single atom transitions where each atom increases its projection of angular momentum by one unit. This simplification of the two-body dynamics allows us to analytically calculate the shifts induced by the interactions on the transition frequencies. These shifts bring the single atom pulse out of resonance therefore disrupting the circularization process. Once the main effect of the interactions between two Rydberg atoms is identified, the correction required by the initial pulse is direct; we propose to adapt the single-atom pulse by modulating it with a phase shift that incorporates the frequency shifts induced by the interactions. Furthermore, we provide a simple functional form for the adapted pulse together with two analytically derived parameters which are also shown to be very close to optimal. 

Our adapted pulses with analytically derived parameters lead to final circular pair state probabilities higher than $\SI{95}{\percent}$ for two $^{87}$Rb atoms in the $n=52$ Stark manifold with interatomic distances above $\SI{6.3}{\micro\meter}$ and for any angle between the interatomic axis and the direction of the static electric and magnetic fields. Moreover, the adapted pulses perform comparably to the ones obtained after a pulse shaping optimization using the numerical Krotov's method with enforced experimental amplitude constraints. The difference in the final circular pair state probability between the adapted pulse with analytical parameters and Krotov's optimization is below $\SI{3}{\percent}$ for interatomic distances above $\SI{5}{\micro\meter}$ and for any angular configuration. Overall we show that the constraints imposed over the algorithm during pulse shaping are the limiting factor on the state preparation, when lifting such constrains we reach our target fidelity of $\SI{99}{\percent}$ for any arrangement with interatomic distances above $\SI{5.9}{\micro\meter}$. Even though the proposed fast RF-field adapted pulses provide for accurate and sufficient controllability to counteract the effects of interactions, we also propose a multilayer optimization scheme that reduces the computational cost: the two linear parameters of the adapted pulse can be optimized and the obtained pulse can be used as initial guess for Krotov's algorithm. It is widely known that pulse shaping algorithms rely on the availability of good initial guess pulses for an efficient optimization. By providing for good initial guess pulses with high performance even for strong interactions or long durations, our proposal contributes to expand the limits of the applicability of quantum control algorithms to the development of circularization protocols while at the same time highlighting the importance of the dialectic between the quantum control machinery and physical insight.

The adapted pulses present several key advantages. Besides the simplicity of the approach, the adapted pulses take into account the effect of quantum defects and comply with experimental constraints by construction, including amplitude constraints which are not implemented in Krotov's algorithm and must therefore be imposed on each iteration with the consequent disruption of the convergence. They also allow for a black box optimization; the two linear parameters can be further optimized in a closed feedback loop by an optimizer that is directly connected to the experiment. In this way the optimizer automatically incorporates experimental effects that are absent in the model, such as transfer functions imposed by the hardware or uncertainties in the static electric and magnetic fields or the interatomic distance and angle between the interatomic axis and the static fields. 

A natural next step is to test our theoretically developed pulses experimentally. Beyond this, it is essential to investigate the circularization of more complex arrangements involving multiple atoms. Due to their exponentially large state space, computing the time evolution for systems with more than two atoms becomes extremely challenging thus rendering the implementation of optimization algorithms unfeasible. However, our proposed adaptation of the single-atom pulse when combined with a closed-loop optimization may enable a realization of fast and accurate pulses even for such demanding settings.

\begin{acknowledgments}
We acknowledge funding from the Horizon Europe program HORIZON-CL4-2022-QUANTUM-02-SGA via the project \href{https://doi.org/10.3030/101113690}{101113690} (PASQuanS2.1) and by Germany’s Excellence Strategy - the Cluster of Excellence Matter and Light for Quantum Computing (ML4Q2) EXC 2004/2 –
390534769. \\
This project has received funding by the French National Research Agency (Grant No. ANR-22-PETQ-004, project QubitAF ; Grant No. ANR-25-CE47-7422-01, project DuRyQS). It has been supported by the Île-de-France region in the framework of DIM QuanTiP (project LT-CRAQS). \\
E. Cuestas was also supported by JSPS KAKENHI grant number JP23K13035.\\
\\
We thank Sebastian Weber for insightful discussions on interacting Rydberg atoms and for assistance with the software package \texttt{pairinteraction} \cite{Weber2017}.\\
\\
During the preparation of this work, the authors used the generative AI tools GitHub Copilot and ChatGPT for code debugging, code optimization, and to improve the readability of written text. The authors validated all generated content and take full responsibility for the content of the manuscript.
\end{acknowledgments}

\begin{appendix}
\label{Appendix}

\section{Details on the simulation}
\label{Appendix: Details on simulation}
\renewcommand{\theequation}{A.\arabic{equation}}
\renewcommand{\thefigure}{A.\arabic{figure}}
\setcounter{figure}{0}

In order to simulate the circularization of a single $^{87}$Rb atom we follow Refs.~\cite{Patsch2018, Patsch2022} and begin by constructing the static part of the system Hamiltonian of Eq.~\ref{Eq.: total single atom Hamiltonian}
\begin{align}
\label{Eq.: static part single atom hamiltonian}
\hat{h}_\text{s} = \hat{h}_0 -\hat{{\mathbf{d}}} \cdot \mathbfcal{E} + \mu_{B}\mathbf{B} \cdot \hat{\mathbf{L}}.
\end{align}
We express all operators in the spherical basis and truncate the Hilbert space as described in Appendix~\ref{Appendix: state space reduction}. Here, $\mu_B$ is the Bohr magneton, $\hat{\mathbf{d}} = -e\, \hat{\mathbf{r}}$ is the dipole operator, with $e$ the electron charge. While most of the eigenstates of the field-free hamiltonian $\hat{h}_0$ are hydrogenic, states with low orbital angular momentum $l$ are affected by quantum defects \cite{gallagher_1994_book, signoles:tel-01146049}, which represent corrections to the corresponding eigenenergies. The matrix elements of the position operator $\hat{{\mathbf{r}}}$ decompose into an angular part, evaluated analytically, and radial part which we compute numerically using Numerov's method \cite{Noumerov1924, gallagher_1994_book, ibali2017}. The angular momentum operator $\hat{\mathbf{L}}$ is diagonal in the spherical basis. In this work we neglect spin-orbit coupling and fix the spin projection to $m_s = 1/2$ wherever necessary. The eigenstates of $\hat{h}_\text{s}$ (see Fig. \ref{fig_single_atom}(a) for the associated eigenenergies) constitute the basis in which we compute time evolutions that is calculated via numerical  diagonalization of $\hat{h}_\text{s}$. For a given RF pulse $\mathbf{F}(t)$ we compute the time evolution by solving the time-dependent Schrödinger equation in the laboratory frame using Chebychev Propagators \cite{Kosloff1994}, as implemented in the \texttt{QuantumPropagators.jl} package \cite{QuantumControl}, with $\hat{h}(t) = \hat{h}_\text{s} - \hat{\mathbf{d}} \cdot \mathbf{F}(t)$.

To simulate the time evolution of a pair of interacting atoms, we construct the Hilbert space as the tensor product of the single-atom eigenbases introduced above. The resulting pair basis states are product states of the form $\ket{\psi_1,\psi_2}  = \ket{\psi_1} \otimes \ket{\psi_2} $ where $\ket*{\psi_{1(2)}}$ denote single-atom eigenstates of $\hat{h}_\text{s}$. In this representation, all single-atom operators act on the respective atomic subspace, i.e., $\hat{H}_0 = \hat{h}_{0} \otimes \mathbb{I} + \mathbb{I} \otimes \hat{h}_0$ (assuming operators \textit{on the left} acting on particle 1 and operators \textit{on the right} acting on particle 2) and equivalently for the terms $\hat{H}_{\mathcal{E}}$, $\hat{H}_{B}$ and $\hat{H}_{\text{RF}}(t)$ in the hamiltonian $\hat{H}(t)$ of Eq.~\ref{Eq.: total atom pair Hamiltonian}. The interaction Hamiltonian of Eq.~\ref{eq: Interaction_hamiltonian} is invariant under particle permutation, therefore the dynamics preserves permutation symmetry. We then consider a symmetrized basis of pair states,
\begin{align}
\ket{\psi_1 , \psi_2}_{\pm} = \frac{\left( \ket{\psi_1 ,\psi_2} \pm \ket{\psi_2, \psi_1} \right)}{\sqrt{2(1 + \delta_{\psi_1 ,\psi_2})}} \,,
\end{align}
so that the symmetric and antisymmetric subspaces are not coupled by the evolution under $\hat{H}(t)$\cite{Weber2017}. Since we only consider evolutions for the atom pair that start in a symmetric state (both atoms occupying the $m=2$ state of the lowest diagonal ladder) we can neglect the antisymmetric subspace of states therefore reducing the state space dimension by a factor of roughly $1/2$.

\section{Guess pulse}
\label{Appendix: Guess Pulse}
\renewcommand{\theequation}{B.\arabic{equation}}
\renewcommand{\thefigure}{B.\arabic{figure}}
\setcounter{figure}{0}

\begin{figure}[tb] 
    \centering
\includegraphics[width=0.49\textwidth]{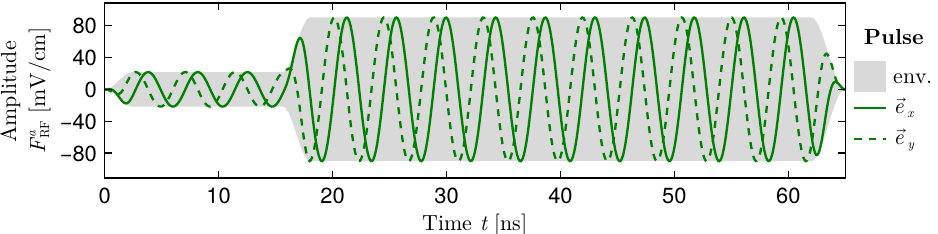} 
    \caption{$x$- and $y$-components (solid and dashed) together with the corresponding envelope (gray shaded area) of the pulse used as initial guess for the optimization of the circularization of a single Rubidium atom. The total duration of the pulse is $t_{\text{f}}=\SI{65}{\nano\second}$ and it satisfies the experimental maximum amplitude constrain of $\SI{92}{\milli\volt\per\centi\meter}$. The pulse amplitude is lowered for initial times to decrease leaking towards the state $\ket{m=1}$.}
    \label{fig: guess pulse} 
\end{figure}

The energy ladders in the fixed-$n$ Stark manifold of a hydrogenic atom are harmonic up to the first order Stark shift. Therefore, for hydrogen a perfect circularization can be achieved by a circular $\sigma_+$ polarized RF pulse. The corresponding evolution can be understood as a $\pi$-rotation on a generalized Bloch sphere \cite{shore_book_1990, Patsch2018, Signoles2017, Facon2016}, meaning that the duration and intensity of the pulse must match the conditions of a $\pi$-pulse (given a maximal experimentally feasible Rabi frequency $\Omega_R$ the amplitude of the pulse is given by $\Omega_R = 3 n e a_0 F_\text{RF}^\text{max}/(2\hbar)$, see Ref.~\cite{Kruckenhauser2022}). In general we aim for the fastest possible pulses. However, for shorter durations the spectral bandwidth of the RF pulse increases and higher intensities are needed. Since the intensity of the pulse must obey the experimental maximum intensity constraint, we use a total duration $t_{\text{f}} = \SI{65}{\nano\second}$ together with a flat-top pulse with sine-squared ramps for initial and final times \cite{Patsch2018, Patsch2022}. This pulse duration leads to a speed-up of the circularization process compared to the currently used adiabatic passage by a factor of approximately $60$ \cite{mehai2023}. The final circular state probability obtained using the simple pulse described above (based on the hydrogenic picture) for a Rubidium atom starting in the $m=2$ state is $\SI{54.4}{\percent}$. This low accuracy can be attributed to quantum defects, that affect low-$m$ states and cause anharmonicities at the beginning of the lowest diagonal ladder. Also, a significant part of the initial population leaks towards the $m=1$ state such that it does not reach the circular state. This leaking mechanism was already identified in Ref.~\cite{Patsch2018} and occurs because the transition from $m = 2$ to $m = 1$ has a small detuning (of $\SI{41.7}{\mega\hertz}$) with respect to the RF pulse, meaning that it is not off-resonant enough to avoid unwanted leakage. To accurately circularize the atom we resort to pulse shaping strategies provided by the quantum control toolbox in order to find more complex pulses that are able to handle the anharmonicities of the non-hydrogenic part (lowest $m$ states) of the lowest diagonal ladder. However, for any pulse optimization routine, a good initial guess pulse already leading to an acceptable final circular state probability $p_\text{C}$ is crucial for a successful and efficient optimization. With this goal in mind and in order to minimize the initial leakage to the $m=1$ state we follow Ref. \cite{Patsch2018} and lower the pulse intensity during the initial stage. The corresponding initial guess pulse used for the single-atom case is shown in Fig.~\ref{fig: guess pulse} for a maximum amplitude of $\SI{90}{\milli\volt\per\centi\meter}$ and total time duration $t_{\text{f}}=\SI{65}{\nano\second}$. During the first $\SI{2.5}{\nano\second}$, the intensity is ramped up to $\SI{21.6}{\milli\volt\per\centi\meter}$ and kept constant for $\SI{13.1}{\nano\second}$. Another ramp of $\SI{2.5}{\nano\second}$ brings the pulse towards a flat-top region with the maximal amplitude of $\SI{90}{\milli\volt\per\centi\meter}$. At the end a $\SI{3}{\nano\second}$ ramp turns the pulse off. All ramps follow a sine-square shape. For static electric and magnetic fields with magnitudes $ \mathcal{E} = \SI{2.1}{\volt\per\centi\meter}$ and $ B = 14 \text{ G}$, we choose a pulse with frequency $\omega_{\text{RF}} = \SI{229.5}{\mega\hertz}$ associated to the average $m \to m+1$ transition energy in the lowest diagonal ladder. The $x$- and $y$-components oscillate with a phase difference of $\varphi_{\text{RF}} = \pi/2$ corresponding to a circular $\sigma_+$ polarization. This guess pulse leads to a final circular state probability of $p_{\text{C}} = \SI{90.8}{\percent}$ for a single Rubidium atom. Finally, it is worth to note that for all the pulses considered in this work, we choose a discretization on time intervals of $\Delta t = 0.1 \ \text{ns}$.

\section{Krotov's method}
\label{Appendix: Krotov's method}
\renewcommand{\theequation}{C.\arabic{equation}}
\renewcommand{\thefigure}{C.\arabic{figure}}
\setcounter{figure}{0}

Originally developed in the context of classical optimal-control theory \cite{Krotov1983}, Krotov's method has been successfully adapted to quantum mechanics particularly for the state preparation problem \cite{Sklarz2002, Reich2012}. It is a gradient-based optimization algorithm that we use to optimize the $x$- and $y$-components of the RF control field $\mathbf{F}(t)$ over multiple iterations $i$ so that the obtained state after the full evolution of the system is a certain predefined target state $\ket{\psi_{\text{tgt}}}$. In our case the target state is the circular state $\ket{n \text{C}}$ for a single-atom system and the circular pair state $\ket{n \text{C}, n \text{C}}$ for an atom pair. The distance to this target is quantified by the infidelity
\begin{align}
    \mathcal{I} = 1 - \abs{\braket{\psi(t_f)}{\psi_{\text{tgt}}}}^2 \, .
\end{align}
We extend this cost function by adding terms that penalize surpassing control constraints \cite{mueller_2022_rep_progress_phys},
\begin{align}
\label{eq_cost_funct}
    J^i = 1 - \abs{\braket{\psi^{i}(t_f)}{\psi_{\text{tgt}}}}^2 + \int_0^{t_f} dt \ g(\mathbf{F}^i(t)),
\end{align}
commonly chosen as 
\begin{align}
\label{eq_cost_func_penalty}
    g(\mathbf{F}^{i}(t)) = \frac{\lambda}{S(t)} \left( \Delta F_x^{i}(t)^2 + \Delta F_y^{i}(t)^2 \right),
\end{align}
where $\Delta F_{x(y)}^{i}(t) = F_{x(y)}^{i}(t) - F_{x(y)}^{i-1}(t)$ is the control update after the $i$-th iteration for each component \cite{Goerz2019}. By tuning the inverse step size $\lambda > 0$ we define how strongly the algorithm changes the controls between consecutive iterations. For our optimizations we set $\lambda = 10^3$. This choice depends on a scaling factor $s = 10^{12}$ with which we re-scaled the energies $s \hat{H}$ and times $t/s$ for numerical stability as we use atomic units in our simulation. We also use a global scaling function $S(t)$ that on each iteration prohibits any pulse update at a time step $t$ by setting $S(t) = 0$. We choose it to have a flat-top shape with sine-square ramps of duration $0.03 \ t_{\text{f}}$ for initial and final times. Consequently, the optimizer does not change the initial guess pulse at the beginning and end, such that the pulses on each iteration have vanishing initial and final amplitude as the guess pulse. The cost functional $J^i$ maps a time evolution $\ket{\psi^i(t)}$ generated by the controls $\mathbf{F}^i(t)$ in the $i$-th iteration to a cost value that we aim to decrease in the subsequent iteration $i+1$. As explained in detail in Refs.~\cite{Goerz2019, Reich2012, Palao2003} the method provides an equation for the control update $\Delta \mathbf{F}^i(t)$, which is computed by sequentially matching a forward propagation of the initial state with a backward propagation of the target state \cite{Patsch2018}. The pulse after iteration $i$ is given as $\mathbf{F}^i(t) = \mathbf{F}^{i-1}(t) + \Delta \mathbf{F}^i(t)$ and guarantees by construction the convergence of the cost functional, $J^{i+1} < J^i$ \cite{Palao2003, Goerz2019}. It is important to highlight that the convergence may be slowed down or completely hindered if constraints on the controls are enforced by post-processing after each iteration (see Appendix~\ref{Appendix: experimental constraints}). 

In this work we employ Krotov's method as implemented in the software package \texttt{QuantumControl.jl} \cite{QuantumControl}. We stop the algorithm if i) it reaches a desired target fidelity $J = 0.99$ (further optimization is omitted at this point since the convergence is slow and it only yields marginal gains that are small compared to the expected experimental errors of around $\SI{3}{\percent}$), ii) the convergence is hindered meaning that we get $J^{i+1} > J^{i}$, or iii) the algorithm converges by satisfying $J^{i} - J^{i+1} \leq 10^{-6}$. 

\section{Implementation of experimental constraints}
\label{Appendix: experimental constraints}
\renewcommand{\theequation}{D.\arabic{equation}}
\renewcommand{\thefigure}{D.\arabic{figure}}
\setcounter{figure}{0}

\begin{figure*}[t]
 \centering
 \includegraphics[width=\textwidth]{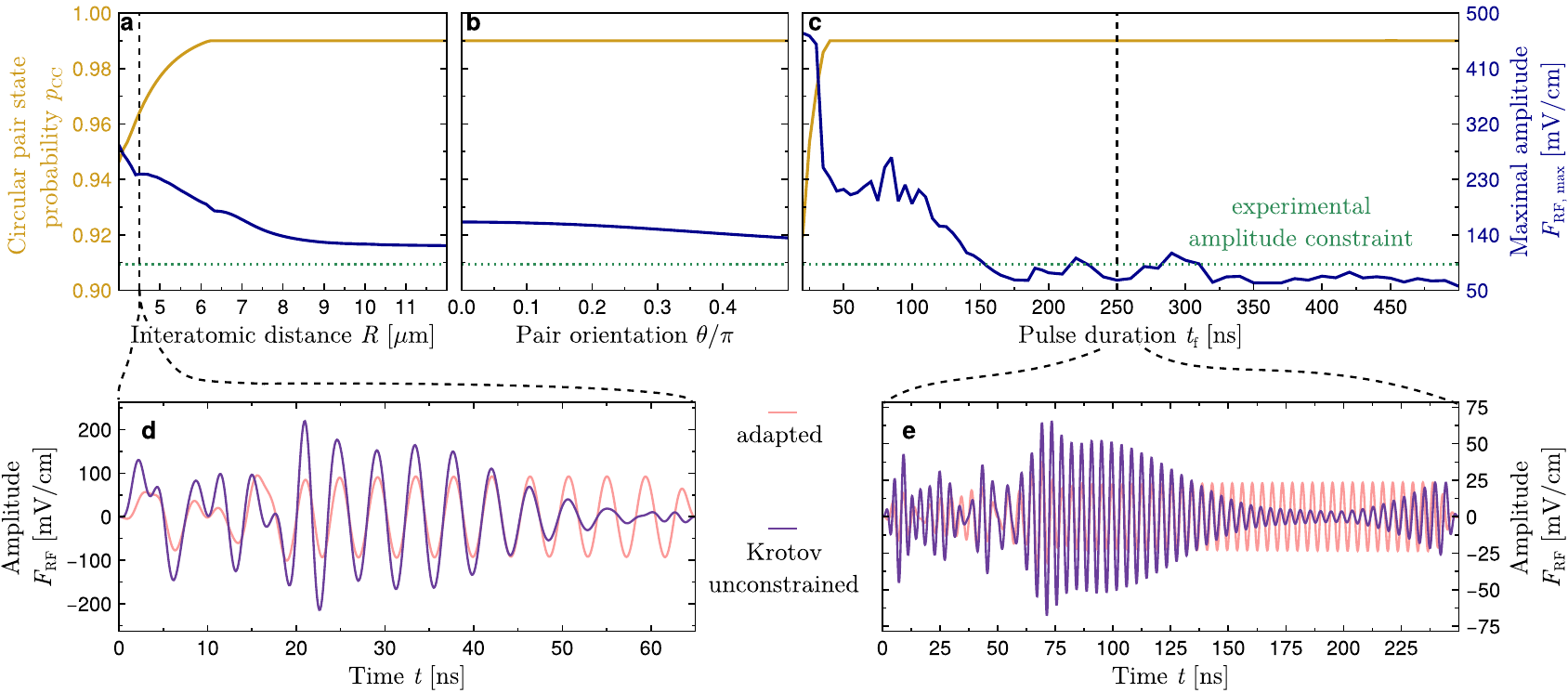}
   \caption{(a-c) Final circular pair state probability $p_\text{CC}$ and maximum of $x$- and $y$-components of the pulses obtained by the unconstrained optimizations with Krotov's method shown in Fig. \ref{fig_comparison_approaches}. The targeted fidelity of $p_{\text{CC}} = \SI{99}{\percent}$ is reached for most interatomic distances $R$, angles $\theta$, and pulse durations $t_{\text{f}}$, but requires pulses with large maximal amplitudes, exceeding the experimental amplitude constraint of $\SI{92}{\milli \volt \per \centi \meter}$ (dotted green line). (d-e) $x$-components of the adapted pulses (based on optimized parameters $\alpha_\text{int}$ and $\beta_{\text{int}}$) and pulses further optimized with Krotov's method without enforcing experimental constraints for $R = \SI{4.5}{\micro \meter}$, $\theta = 0$ and $t_{\text{f}} = \SI{65}{\nano \second}$, and $R = \SI{7}{\micro \meter}$, $\theta = 0$ and $t_{\text{f}} = \SI{250}{\nano \second }$, respectively.}
 \label{fig_max_amps}
\end{figure*}

To ensure the experimental feasibility of the obtained pulses, we must implement the constraints imposed by the experimental hardware such as limits in the frequencies and intensity of the fields. In principle, we could design a corresponding penalty term $g$ and include it in the cost functional $J$, see Eqs.~\eqref{eq_cost_funct} and \eqref{eq_cost_func_penalty} of Appendix \ref{Appendix: Krotov's method}. However, to the best of our knowledge, there exists no implementation of Krotov's method that inherently incorporates arbitrary constraint terms $g$ \cite{Goerz2019}. To overcome this issue we follow Ref.~\cite{Patsch2018} and enforce the constraints by post-processing the pulses after each iteration. This extra step can lead to a slower convergence and, when the pulse update is strongly altered, the convergence can be completely disrupted. 

We post-process our obtained pulses as follows: i) we limit $x$ and $y$-amplitudes to a maximal value $F_\text{RF}^\text{max} = \SI{92}{\milli\volt\per\centi\meter}$ respectively based on the $\pi-$pulse condition and fixing the maximal experimentally feasible Rabi frequency as $\Omega_R/{(2\pi)} = \SI{9}{\mega\hertz}$~\cite{Patsch2018}, ii) we enforce the frequency bandwidth constraints by setting all components above $\SI{480}{\mega\hertz}$ to zero. iii) Since step ii) can lead to non-zero amplitudes at the beginning and end of the pulse, we frame the pulse with a scaling function $S(t)$ setting all the pulse values satisfying $F_{x(y)}(t) > S(t)$ to $S(t)$. We compare three scenarios: experimental constraints (i.e. $0-\SI{480}{\mega\hertz}$ and $\SI{92}{\milli\volt\per\centi\meter}$), relaxed amplitude constraints (i.e. $0-\SI{480}{\mega\hertz}$ and $\SI{130}{\milli\volt\per\centi\meter}$), and finally unconstrained meaning no constraints at all.

Experimental amplitude constraints are the major limitation for Krotov's optimization, leading to slower convergence and lower final circular pair state probabilities compared to the unconstrained case, see Fig. \ref{fig_comparison_approaches} in the main text. If the experimental constraints are lifted, the optimization reaches the targeted fidelity of $p_{\text{CC}} = \SI{99}{\percent}$ for interatomic distances $R \geq \SI{5.9}{\micro \meter}$ and a pulse duration of $t_{\text{f}} = \SI{65}{\nano \second}$ or for $t_{\text{f}} \geq \SI{40}{\nano \second}$ and $ R=\SI{7}{\micro \meter}$. However, as shown in Fig. \ref{fig_max_amps}, this comes at the costs of amplitudes exceeding the experimental amplitude constraint even by a factor of five.

\section{Couplings induced by Interactions}
\label{Appendix:couplings induced by interactions}
\renewcommand{\theequation}{E.\arabic{equation}}
\renewcommand{\thefigure}{E.\arabic{figure}}
\renewcommand{\thetable}{E.\Roman{table}}
\setcounter{figure}{0}
\setcounter{table}{0}

In what follows we aim to show that the coupling induced by interactions among pair states in the lowest diagonal ladder to pair states in upper diagonal ladders is weak enough to be neglected. We assume fixed electric and magnetic field strength, $\mathcal{E}$ and $B$; interatomic distance $R$; and angle $\theta$ between the interatomic axis and the direction of the static fields. Given the symmetry of the Coulomb potential, the states belonging to the bound part of the spectrum of an hydrogenic atom can be represented by two coupled angular momenta and the dipole-dipole interaction can be understood as a coupling of four angular momenta \cite{englefield_book_1972,galindo_springer_1990,shore_book_1990, Kruckenhauser2022}. For a single atom, two linear combinations of the usual orbital angular momentum $\hat{\bf{L}}$ and the Runge-Lenz vector $\hat{\bf{K}}$ can be constructed as $\hat{\bf{J}}^{\text{p(m)}} = (\hat{\bf{L}} \pm \sqrt{-m/(2 \hat{h}_0)} \hat{\bf{K}})/2$, where $\hat{h}_0$ denote the hydrogenic hamiltonian $\hat{h}_0 = - m e^4/ (2(2((\hat{\bf{J}}^\text{p})^2 + (\hat{\bf{J}}^\text{m})^2)+\hbar^2))$ where the superscripts refer to the plus (p) minus (m) combinations. Both $\hat{\bf{J}}^{\text{p(m)}}$ satisfy the angular momentum commutation relationships and commute with each other, therefore, the usual angular momentum lowering and raising operators $\hat{J}^{\text{p(m)}}_\pm = \hat{J}^{\text{p(m)}}_x \pm i \hat{J}^{\text{p(m)}}_y $ give rise to the states $\ket{j, m_\text{m}, m_\text{p}} \equiv \ket{m_\text{m}, m_\text{p}}$ where $m_\text{p(m)}=-j, -j + 1, ..., j$ with $j = (n-1)/2$ \cite{shore_book_1990, Kruckenhauser2022}. It is important to keep in mind that because of the definition of $\hat{\bf{J}}^\text{p(m)}$ all what follows is valid within a fixed-$n$ manifold, see also \cite{englefield_book_1972,galindo_springer_1990}. The quantum numbers associated with the eigenvalues of $\hat{J}^\text{p(m)}_z$, i.e. $m_\text{p(m)}$, are related to the ones of $\hat{L}_z$ and $\hat{K}_z$ through $m_l=m_\text{p}+m_\text{m} \equiv m$ and $m_k=m_\text{p}-m_\text{m} \equiv k$. Remember that $k$ is the eccentricity quantum number that is related to the quantum numbers used in the parabolic basis definitions (for example, to the $n_1$ and $n_2$ of Refs.~\cite{Bethe1957, Gallagher1988} as $k=n_1-n_2$ with $n_1+n_2=n-|m|-1$), see also Refs.~\cite{jannussis_pla_1979,lai_pla_1981} for the perturbative calculation of higher order shifts due to the Stark effect. The circular states are then the states with the extremal spin orientations $m_\text{p}=m_\text{m}=\pm j$, or equivalently, $k=0$ and $m=\pm(n-1)$; the lowest diagonal ladder corresponds to the states with $m_\text{m}=j$ and $m_\text{p}=-j, \dots, j$ or $k=-n+1+m$ and $m=0,\dots,n-1$; generically, the $i$-th diagonal ladder is given by $m_\text{m} = j-(i-1)$ and $m_\text{p}=-j, \dots, j$ or $k=-n+(2i-1)+m$ and $m=-(i-1),\dots,n-1-(i-1)$. All of this means that the intra-ladder transitions are given by the operators $\hat{J}^\text{p}_{\pm}$ (\textit{up} and \textit{down} the ladder respectively), while the inter-ladder couplings are associated with the operators $\hat{J}^\text{m}_{\pm}$ (to a lower or upper diagonal ladder, respectively). 

\begin{table*}[t]
    \centering
    \[
    \begin{array}{ccccc}
      \text{Term} & \ket{\Psi'}  & \max\limits_{(\theta, m_\text{p})}
      \left(\abs{\bra{\Psi'}\hat{H}_{\text{int}}\ket{\Psi}}\right) & \Delta E  & R_{\text{min}} [\SI{}{\micro\meter}] \\ \hline 
      \hat{J}^{\text{m},1}_- \hat{J}^{\text{p},2}_+ + \hat{J}^{\text{p},1}_+ \hat{J}^{\text{m},2}_- 
         & \frac{\ket{j-1, m_\text{p}; j, m_\text{p}+1}+\ket{j, m_\text{p}+1 ; j-1, m_\text{p}}}{\sqrt{2}}
         & \frac{9e^2 a_0^2}{32 \sqrt{2} \pi \epsilon_0 R^3} n^3 \sqrt{n-1}  
         & 3 e a_0 n\mathcal{E} 
         & 1.2 \\[2mm]
      \hat{J}^{\text{m},1}_- \hat{J}^{\text{p},2}_- + \hat{J}^{\text{p},1}_- \hat{J}^{\text{m},2}_- 
         & \frac{\ket{j-1, m_\text{p} ; j, m_\text{p}-1}+\ket{j, m_\text{p}-1 ; j-1, m_\text{p}}}{\sqrt{2}}
         & \frac{27 e^2 a_0^2}{64\sqrt{2} \pi \epsilon_0 R^3} n^3 \sqrt{n-1} 
         & -2\mu_B B 
         & 3.1 \\[2mm]
      \hat{J}^{\text{m},1}_- \hat{J}^{\text{m},2}_- 
         & \ket{j-1, m_\text{p};j-1, m_\text{p}}
         & \frac{27e^2 a_0^2}{64 \pi \epsilon_0 R^3} n^2 (n-1) 
         & 3 e a_0 n\mathcal{E}-2\mu_B B 
         & 0.8 \\[2mm]
      \hat{J}^{\text{m},1}_- \hat{J}^{\text{p(m)},2}_z + \hat{J}^{\text{p(m)},1}_z \hat{J}^{\text{m},2}_- 
         & \frac{\ket{j-1, m_\text{p}; j, m_\text{p}}+\ket{j, m_\text{p}; j-1, m_\text{p}}}{\sqrt{2}} 
         & \frac{27 \sqrt{2} e^2 a_0^2}{128 \pi \epsilon_0 R^3} n^2 (n-1) \sqrt{n-1} 
         & \frac{3}{2} e a_0 n\mathcal{E} - \mu_B B 
         & 1.8
    \end{array}
    \]
    \caption{Maximal matrix elements (maximised over all the possible angles $\theta$ and lowest diagonal ladder states $m_\text{p}=-j,\dots,j$) for each term contributing to the interaction Hamiltonian explicitly written in Eq.~\eqref{eq: interaction hamiltonian parabolic basis expanded} and the corresponding energy gaps computed between pair states in the lowest and second lowest diagonal ladder. For the chosen static field strengths $\mathcal{E}= \SI{2.1}{\volt\per\centi\meter}$ and $B = 14 \text{ G}$ we check the perturbative condition of Eq.~\eqref{eq: small pertubation condition} for each relevant term in the interaction Hamiltonian and derive a minimal interatomic distance $R_{\text{min}}$ such that the associated term can be neglected whenever $R \gg  R_{\text{min}}$.}
    \label{tab: all interaction terms conditions}
\end{table*}

Inside a fixed-$n$ manifold and up to the first order Stark shift, the coupling between an hydrogenic atom and a  static electric field $\mathbf{\mathcal{E}}$ and magnetic field $\mathbf{B}$ can be written as   
\begin{align}
	\hat{h}_{\mathcal{E}} &+ \hat{h}_{\text{B}} = \notag \\
	&-  \left( \frac{3}{2}e a_0 n \mathcal{E} - \mu_B B \right) \hat{J}^\text{m}_z + \left( \frac{3}{2}e a_0 n \mathcal{E} + \mu_B B \right) \hat{J}^\text{p}_z \,, \notag
\end{align}
see Ref.~\cite{Kruckenhauser2022} for a discussion on the conditions for this last expression to hold (Ingris-Teller limit and negligible diamagnetic couplings). For the sake of completeness we would like to mention that the expression for the Hamiltonian in the presence of the time-dependent circularly polarized RF-field can also be greatly simplified in the rotating frame and after the rotating wave approximation (see also Ref.~\cite{Kruckenhauser2022} together with the beautiful and exhaustive book of Bruce W. Shore Ref.~\cite{shore_book_1990}) leading to an $n$-level Rabi system for the lowest diagonal ladder that depicts two-states periodicity setting the basis for the adiabatic passage circularization technique (see Refs.~\cite{hulet_1983,rubbmark_pra_1981,Signoles2017}) as well as the proposal of creating an spin coherent state for its subsequent rotation exploited in Refs.~\cite{Patsch2018,Larrouy2020}. Note that our numerical calculations are performed in the laboratory frame and we do not rely in the rotating wave approximation.

To derive the regime or conditions under which the couplings induced by interactions can be neglected, we assume weak interactions and treat them as a perturbation to $\hat{H}_\text{s} = \hat{H}_0 + \hat{H}_{\mathcal{E}} + \hat{H}_{\text{B}} = \mathbf{1} \otimes \hat{h}_\text{s} + \hat{h}_\text{s} \otimes \mathbf{1}$, with $\hat{h}_\text{s} = \hat{h}_0 + \hat{h}_{\mathcal{E}} + \hat{h}_{\text{B}}$. As also explained in Ref.~\cite{Kruckenhauser2022} the dipole-dipole interactions can be written as
\begin{align}
\label{eq: interaction Hamiltonian parabolic basis}
        \hat{H}_{\text{int}} = V 
         \Big[& \hat{\mathbf{J}}^{\text{m},1} \cdot \hat{\mathbf{J}}^{\text{m},2} 
        - 3 (\hat{\mathbf{J}}^{\text{m},1} \cdot \mathbf{e}_R) (\hat{\mathbf{J}}^{\text{m},2} \cdot \mathbf{e}_R) \notag \\
         + & \hat{\mathbf{J}}^{\text{p},1} \ \cdot \hat{\mathbf{J}}^{\text{p},2} \ 
        - 3 (\hat{\mathbf{J}}^{\text{p},1} \ \cdot \mathbf{e}_R) (\hat{\mathbf{J}}^{\text{p},2} \ \cdot \mathbf{e}_R) \notag\\
         - & \hat{\mathbf{J}}^{\text{m},1} \cdot \hat{\mathbf{J}}^{\text{p},2} \ 
        + 3 (\hat{\mathbf{J}}^{\text{m},1} \cdot \mathbf{e}_R) (\hat{\mathbf{J}}^{\text{p},2} \ \cdot \mathbf{e}_R) \notag\\
         - & \hat{\mathbf{J}}^{\text{p},1} \ \cdot \hat{\mathbf{J}}^{\text{m},2}
         + 3 (\hat{\mathbf{J}}^{\text{p},1} \ \cdot \mathbf{e}_R) (\hat{\mathbf{J}}^{\text{m},2} \cdot \mathbf{e}_R)
    \Big] \, ,
\end{align}
where $V = \frac{(3e a_0 n)^2}{16\pi \epsilon_0 R^3}$ and $\mathbf{e}_R$ is the unit vector along the interatomic axis $\mathbf{e}_R = (\sin{\theta}, 0, \cos{\theta})$, see Fig. \ref{fig_two_atoms_single_atom_pulse}(a) in the main text. The numbers in the operators refer to atom 1 or 2 within the pair. We assume that the interatomic distance $R$ is sufficiently large for the atoms to be distinguished (i.e. $R>R_{LR}$ with $R_{LR}$ being the Le Roy radius so that the electronic clouds do not overlap \cite{Weber2017}). Using the expressions for $\hat{J}^{\text{p(m)}}_\pm$ the latter equation reads
\begin{align}
\label{eq: interaction hamiltonian parabolic basis expanded}
    \hat{H}_{\text{int}} = 
    \frac{V}{2}  \Big[
    & \left(\frac{3}{2} \sin^2\theta -1 \right) \left( \hat{J}^{\text{m},1}_- \hat{J}^{\text{p},2}_+ + \hat{J}^{\text{p},1}_+ \hat{J}^{\text{m},2}_- \right)   \notag\\ 
    + & \frac{3}{2} \sin^2\theta \left( \hat{J}^{\text{m},1}_- \hat{J}^{\text{p},2}_- + \hat{J}^{\text{p},1}_- \hat{J}^{\text{m},2}_- \right) \notag \\
    - & \frac{3}{2} \sin^2\theta \, \hat{J}^{\text{m},1}_- \hat{J}^{\text{m},2}_- \notag \\
    + & 3\sin\theta \cos\theta \left( \hat{J}^{\text{m},1}_- \hat{J}^{\text{p},2}_z + \hat{J}^{\text{p},1}_z \hat{J}^{\text{m},2}_-  \right) \notag \\
    - & 3\sin\theta \cos\theta \left( \hat{J}^{\text{m},1}_- \hat{J}^{\text{m},2}_z + \hat{J}^{\text{m},1}_z \hat{J}^{\text{m},2}_-  \right) \Big] \notag \\
    + & \dots \, ,
\end{align}
where we only show terms containing $\hat{J}^{\text{m}}_{-}$ that couple an atom in one of the lowest diagonal ladder states $\ket{\psi} = \ket{m_\text{m}=j, m_\text{p}}$ to the next higher ladder $\ket{\psi'} = \ket{ m_\text{m}=j-1, m_\text{p}}$ via
\begin{align}
    \hat{J}^{\text{m}}_{-} \ket{j, m_\text{p}} 
    = & \sqrt{j(j+1) - j(j-1)}\ket{j-1, m_\text{p}} \\
    = & \sqrt{n-1}\ket{j-1, m_\text{p}}. \notag
\end{align}

We focus on a pair state $\ket{\Psi} = \ket{m_\text{m}^1=j, m_\text{p}^1=m_\text{p}; m_\text{m}^2=j, m_\text{p}^2=m_\text{p}}$ where both atoms are in the same lowest diagonal ladder state $\ket{\psi} = \ket{j, m_\text{p}}$ and $M_- =0$. Each term of $\hat{H}_{\text{int}}$ containing the operator $\hat{J}^\text{m}_{-}$ couples $\ket{\Psi}$ to the symmetric pair states $\ket{\Psi'}$ where at least one of the atoms occupies a higher diagonal ladder state $\ket{\psi'}=\ket{j-1, m'_\text{p}}$ with $m'_\text{p} = m_\text{p}$ or $m_\text{p} \pm 1$. This coupling can be neglected as long as the corresponding matrix element is small compared to the energy gap  $\Delta E = \bra{\Psi'}\hat{H}_{\text{s}}\ket{\Psi'} - \bra{\Psi}\hat{H}_{\text{s}}\ket{\Psi}$ between the pair states $\ket{\Psi}$ and $\ket{\Psi'}$ given by the unperturbed Hamiltonian, 
\begin{align}
\label{eq: small pertubation condition}
    \abs{\bra{\Psi'}\hat{H}_{\text{int}}\ket{\Psi}} \ll \abs{\Delta E},
\end{align}
where $\bra{\Psi}\hat{H}_{\text{s}}\ket{\Psi} = 3e a_0 n\mathcal{E} (m_\text{p}-j) -2\mu_B B (m_\text{p}+j)$ since the single particle energy of a state $\ket{\psi}=\ket{m_\text{m}, m_\text{p}}$ associated to $\hat{h}_{\text{s}}$ is $\bra{\psi}\hat{h}_{\text{s}}\ket{\psi} = \frac{3}{2}e a_0 n\mathcal{E} (m_\text{p}-m_\text{m}) -\mu_B B (m_\text{p}+m_\text{m})$.

Now we proceed to check the condition of Eq.~\eqref{eq: small pertubation condition} for all terms in Eq.~\eqref{eq: interaction hamiltonian parabolic basis expanded}. The first term $\propto \hat{J}^{\text{m},1}_-\hat{J}^{\text{p},2}_+ + \hat{J}^{\text{p},1}_+\hat{J}^{\text{m},2}_-$ couples $\ket{\Psi}$ to the symmetrized pair states 
\begin{align}
\ket{\Psi'} = \frac{\ket{j-1, m_\text{p};j, m_\text{p}+1}+\ket{j, m_\text{p}+1;j-1, m_\text{p}}}{\sqrt{2}}
\end{align}
through
\begin{align}
\big|\bra{\Psi'}& \frac{V}{2} \left(\frac{3}{2} \sin^2\theta -1 \right) \left( \hat{J}^{\text{m},1}_- \hat{J}^{\text{p},2}_+ + \hat{J}^{\text{p},1}_+ \hat{J}^{\text{m},2}_- \right)   \ket{\Psi}\big| \notag \\ 
    &= \sqrt{n-1} \bra{j, m_\text{p} + 1} \hat{J}^\text{p}_+ \ket{j, m_\text{p}} \frac{V \vert 1 - \frac{3}{2} \sin^2\theta \vert}{\sqrt{2}}  \notag \\
    & = \sqrt{(n-1) \left( \frac{n^2-1}{4} - m_\text{p}(m_\text{p}+1) \right)} \notag \\
    & \qquad \qquad \qquad \qquad \qquad \qquad  \times \frac{V \vert 1 - \frac{3}{2} \sin^2\theta \vert}{\sqrt{2}}   \notag \\
    & \leq \sqrt{(n-1) \left( \frac{n^2-1}{4} - \left( - \frac{1}{2}\right)\left(- \frac{1}{2}+1\right) \right)} \frac{V}{\sqrt{2}}  \notag \\
    & \leq \sqrt{n-1} \frac{V n }{2 \sqrt{2}},
\end{align}
which we maximised over all the possible angles $\theta$ and lowest diagonal ladder states $m_\text{p} = -j,\dots,j$. The corresponding energy gap is $\Delta E  = 3 e a_0 n\mathcal{E}$. Following the condition of Eq.~\eqref{eq: small pertubation condition}, the couplings can be neglected as long as 
\begin{align}
    R \gg R_{\text{min}} = \left( \sqrt{n-1}n^2 \frac{3 e a_0 }{32 \sqrt{2} \pi \epsilon_0 \mathcal{E}}\right)^{1/3} \, ,
\end{align}
which for the parameters of our setting reduces to $ R_\text{min} \approx \SI{1.2}{\micro\meter}$. 

The corresponding conditions for the remaining terms are summarized in Table ~\ref{tab: all interaction terms conditions}. Most of the terms lead to negligible couplings with the exception of the term $\propto \hat{J}^{\text{m},1}_- \hat{J}^{\text{p},2}_- + \hat{J}^{\text{p},1}_- \hat{J}^{\text{m},2}_- $ that couples pair states whose energy gap is given by the Zeeman-splitting which does not have a dependence on $n$ and therefore can be small. The effect of this dominant term can be reduced by increasing the magnetic field. In particular, a field $50 \text{ G}$ would lead to $R_{\text{min}} \leq \SI{2}{\micro\meter}$ for all terms. 

\section{State space reduction}
\label{Appendix: state space reduction}
\renewcommand{\theequation}{F.\arabic{equation}}
\renewcommand{\thefigure}{F.\arabic{figure}}
\renewcommand{\thetable}{F.\Roman{table}}
\setcounter{figure}{0}
\setcounter{table}{0}

In order to minimize the computational cost associated with our numerical simulations, we reduce the state-space dimension while maintaining a desired precision in the final probability distribution. In what follows we provide details on our state space truncation to simulate the time evolution of the single atom and the atom pair starting from an initial state located in the lowest diagonal ladder of the $n=52$ Stark manifold and under the action of a time dependent circularly polarized RF field with a frequency $\omega_{\text{RF}} \approx \SI{200}{\mega\hertz}$. 

Since the energy associated to the RF frequency is much smaller than the energy gaps with states in the adjacent $n' = n \pm 1, n \pm 2, \dots$ manifolds (approximately $\SI{10}{\giga\hertz}$), we neglect the coupling to states in manifolds with $n \neq 52$ and only include the states with $n=52$ in our simulations for the time evolution. For the calculation of the static Hamiltonian eigenstates, however, we include all the hydrogenic states in the $n=48,\dots,56$ manifolds. The leakage to higher ladders induced by the fast drive pulses ($t_f \approx \SI{65}{\nano\second}$) is reduced by the detuning of the $\sigma_-$ transitions with respect to $\omega_{\text{RF}}$ introduced by the static magnetic field $\mathbf{B}$. 

In all of our simulations for the evolution of a single atom we include states in the five lowest diagonal ladders. The latter leads to a space state dimension of 250 with an associated error of approximately $10^{-3}\si{\percent}$ in the final circular state probability $p_\text{C}$ compared to a simulation with the ten lowest diagonal ladders. This error is smaller than the targeted precision of $10^{-2}\si{\percent}$ and is in turn much smaller than the experimental errors estimated to be as $\SI{3}{\percent}$. Taking into account the small order of the errors it is possible to include fewer ladders in order to speed up the simulation of the evolution of a single atom. 

\begin{figure*}[t] 
    \centering
    \includegraphics[width=\textwidth]{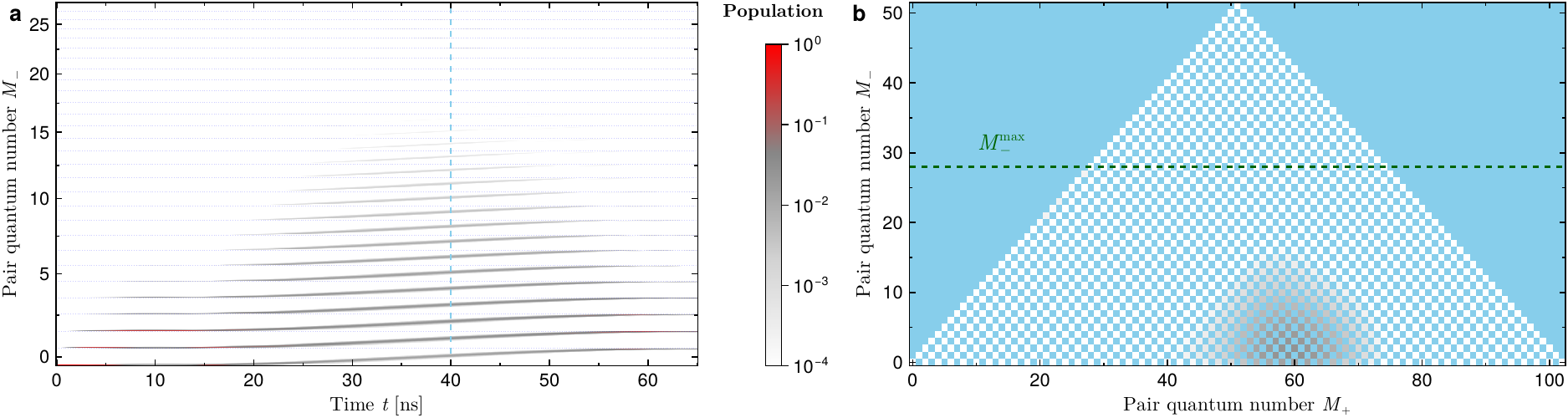}
    \caption{Population distribution over the pair states supported by the lowest diagonal ladders of two interacting Rb atoms with an interatomic distance of $\SI{7}{\micro\meter}$ parallel to the static electric and magnetic field and for a pulse duration $t_\text{f}=\SI{65}{\nano\second}$. (a) Population over the $M_-= m_2 - m_1$ states as a function of time during the evolution driven by the single-atom pulse depicted in Fig.~\ref{fig_single_atom}(b) of the main text. (b) Population distribution over the pair states $\ket{M_+, M_-}$ for a snapshot at $t=\SI{40}{\nano\second}$ of the evolution depicted in panel (a). The vertical blue line in panel (a) indicates the time corresponding to the results of panel (b). At $t=\SI{0}{\nano\second}$ all the population is located in $\ket{M_+ = 4, M_- = 0}$ (bottom left corner) with the goal of evolving it along the bottom edge towards the circular state $\ket{M_+ = 102, M_- = 0}$ (bottom right corner). During the evolution, pair states with large $M_-$ are barely populated and therefore we neglect all the states with $M_- \geq M_-^\text{max} = 28$. This bound introduces an error in the final circular pair state probability $p_\text{CC}$ of approximately $0.01\si{\percent}$. We calculate the truncation value $M_-^\text{max}$ for each simulation.}
    \label{fig: M_pyramid} 
\end{figure*}

For the evolution of the atom pair, we leverage the weak coupling between pair states in the lowest diagonal ladders and pair states in upper diagonal ladders (see Appendix~\ref{Appendix:couplings induced by interactions}) and we consider pair states given by product states among single-particle states belonging to the five lowest diagonal ladders, thus including a total of $250^2 =62 500 $ two-body states. The pair basis can be further reduced by exploiting the permutation symmetry that is preserved by the dipole-dipole interaction, see Ref.~\cite{Weber2017}. In a symmetrized pair basis, the Hamiltonian of Eq.~\eqref{Eq.: total atom pair Hamiltonian} couples pair states with the same symmetry. Since our initial pair state is symmetric under particle exchange, the atom pair evolves in the symmetric subspace meaning that the antisymmetric subspace can be neglected effectively reducing the pair basis dimension by half. The corresponding pair state space on the symmetrized basis has a dimension of 31370 which can be further reduced due to the fact that during the evolution the pair states $\ket{M_+, M_-}$ with a large difference $M_- = m_1 - m_2$ between magnetic quantum numbers $m_1$ and $m_ 2$ are sparsely populated --, see Fig.~\ref{fig: M_pyramid}(a)-- allowing us to implement a cut-off. Figure~\ref{fig: M_pyramid}(b) shows a time snapshot of the population over the pair states supported by the lowest diagonal ladders with the corresponding truncation on $M_-$ due to the very low population present in states with $M_- \geq M_-^\text{max} = 28$. In practice, we find upper bounds on $M_-^\text{max}$ for each ladder combination by systematically lowering an initial bound of $n-1$ until we break our target precision of $0.01 \si{\percent}$ for the final circular pair state probability $p_\text{CC}$. Pair states with $M_- \geq M_-^\text{max}$ are neglected thereupon. The number of pair states included in our simulations are shown in Table~\ref{tab: M- thresholds} for all possible combination of ladders. For each atom pair arrangement $(R, \theta)$ we verify compliance to our target precision by comparing to simulations that include all the pair states supported by the five lowest diagonal ladders. Overall, we reduce the pair basis dimension to a total of 8131 pair states leading to a much more efficient implementation of time evolution simulations.

\begin{table}[h]
    \centering
    \[
    \begin{array}{c|ccccc}
       \text{Ladders} & 1 & 2 & 3 & 4 & 5\\ \hline
      1 & 28 & 26 & 15 & 6 & 3\\
      2 & 26 & 18 & 12 & 8 & 0\\
      3 & 15 & 12 & 3  & 7 & 0\\
      4 & 6  & 8  & 7  & 0 & 0\\
      5 & 3  & 0  & 0  & 0 & 0
    \end{array}
    \]
    \caption{Number of pair states $M_{-}^{\text{max}}$ included in our simulations for two-body states in different diagonal ladders (rows and columns). For example, when having one atom in the second-lowest and the other one in the third-lowest ladder the pair states are neglected if $M_- \geq 12$.}
    \label{tab: M- thresholds}
\end{table}

The obtained maximum cut-off of 28 states is consistent with the bound obtained for the standard deviation $\sigma_m \leq 4$ based on the evolution under the action of the pulse optimized for the circularization of a single $^{87}$Rb atom, see Fig.~\ref{fig_single_atom}(b) of the main text. In the non-interaction case (sufficiently large distance) the distributions for $m_1$ and $m_2$ are independent and therefore the standard deviation for the difference $M_-$ is $\sigma_{M_-} = \sqrt{\sigma_{m_1}^2 + \sigma_{m_2}^2} \leq 5.7$. The bound of 28 states found for the intermediate interaction regime is consistent with the interval $\pm 3 \sigma_{M_-}$ containing 34 states for the non-interacting case, supporting our assumption of a small variance for the $M_-$ distribution that (together with the fact that due to the symmetry of the Hamiltonian $\expval*{\hat{L}^{1}_z-\hat{L}^{2}_z} = 0$ during the evolution) is key for the physical picture of both atoms simultaneously climbing their corresponding lowest diagonal ladder. 

Note that the state space reduction is tailored for the circularization task. For a different state preparation problem (for instance, cat states \cite{Larrouy2020}) the state space reduction must be adjusted accordingly. 

\end{appendix}
\bibliography{ry_circ_bib}
\bibliographystyle{apsrev4-2}

\end{document}